\documentclass{aa}  

\usepackage{txfonts}
\usepackage{graphicx}	
\usepackage{amsmath}	
\usepackage{amssymb}	
\usepackage{comment}
\usepackage{soul}
\usepackage{color}
\usepackage{xcolor}
\usepackage{hyperref}
\usepackage{multirow}

\def\msun{{\rm M_{\odot}}}
\def\rsun{{\rm R_{\odot}}}

\def\msunyr{{\rm M_{\odot}~\rm{yr}^{-1}}}

\def\SN#1{\textcolor{red}{#1}}
\def\RK#1{\textcolor{blue}{#1}}
\def\VE#1{{\bf #1}}
\def\ACG#1{\textcolor{orange}{#1}}
\def\DW#1{\textcolor{magenta}{#1}}

\def\SN#1{{#1}}
\def\RK#1{{#1}}
\def\VE#1{{#1}}
\def\ACG#1{{#1}}
\def\DW#1{{#1}}

\def\REF#1{\textcolor{black}{#1}}
\def\REFF#1{\textcolor{black}{#1}}

\begin{document}

   \title{Episodic accretion in high-mass star \RK{formation}: \RK{An} analysis of thermal instability \RK{for axially symmetric disks}}


   \author{ Vardan G. Elbakyan\inst{1},
            Dennis Wehner\inst{1},
            Rolf Kuiper\inst{1},
            Sergei Nayakshin\inst{2}, 
            Alessio Caratti o Garatti\inst{3}, 
            Zhen Guo\inst{4,5}
          }
   \institute{Fakultät für Physik, Universität Duisburg-Essen, Lotharstraße 1, D-47057 Duisburg, Germany; 
   \email{vardan.elbakyan@uni-due.de} 
    \and
     School of Physics, University of Leicester, Leicester, LE1 7RH, UK;  
    \and
    INAF-Osservatorio Astronomico di Capodimonte, Salita Moiariello 16, 80131 Napoli, Italy;
    \and
    Instituto de Física y Astronomía, Universidad de Valparaíso, ave. Gran Bretaña, 1111, Casilla 5030, Valparaíso, Chile;
    \and
    Millennium Institute of Astrophysics, Nuncio Monse{\~n}or Sotero Sanz 100, Of. 104, Providencia, Santiago, Chile
}

    \titlerunning{2D TI in HMYSOs}
    \authorrunning{Elbakyan et al.}

   \date{Received March 15, 2025; accepted May 15, 2025}

 
  \abstract
   {\ACG{Similarly to their low-mass counterparts, h}igh-mass young stellar objects exhibit episodic accretion bursts. Understanding the physical mechanisms behind these bursts is crucial for elucidating the early stages of massive star formation \RK{and the evolution of disks around high-mass protostars}.}
   {This study aims to investigate the role of thermal instability in triggering accretion outbursts by developing a two-dimensional hydrodynamical model that fully resolves the vertical structure of the inner disk. Our goal is to provide a more realistic depiction of \RK{axially symmetric} disk dynamics during these events and to assess the observable signatures of such bursts.}
   {We perform simulations of the inner 10 astronomical units of a circumstellar disk surrounding a high-mass protostar. The model incorporates heating from viscous dissipation and radiative transport in both the radial and vertical directions. 
   \SN{Unlike previous one-dimensional studies, our two-dimensional \RK{axially symmetric} study resolves the \RK{time-dependent} vertical disk structure, capturing complex} interplay between radial and vertical dynamics within the disk.}
   {Our simulations reveal that thermal instability leads to significant changes in the disk structure. In the inner regions, steep temperature gradients and vigorous convective motions \SN{develop at the onset of outbursts, with gas flows differing between the midplane and the upper disk layers rather than following a purely one-dimensional pattern}. The energy released during the burst is distributed gradually throughout the disk, producing outbursts with durations of 15–30 years and peak mass accretion rates in the range of $2-3\times10^{-4}~\msunyr$. Although these bursts 
   \SN{are observable, they are insufficiently bright, and} their rise times and overall profiles differ from some of the more rapid events seen in observations. \SN{Notably, our models also do not produce the \RK{weaker} “reflares” that sometimes occur atop \RK{stronger} outbursts in one-dimensional thermal instability calculations.}}
   {Resolving the full vertical structure of the disk is essential for accurately modeling thermal instability outbursts in high-mass young stellar objects. While thermal instability significantly influences episodic accretion, our results suggest that it appears insufficient on its own to explain the full range of observed outburst phenomena in HMYSOs. Additional mechanisms \RK{seem to} be required to fully explain the diversity of observed burst phenomena. Future studies incorporating further physical processes are needed to develop a comprehensive understanding of episodic accretion in massive star formation.}


   \maketitle
%


\section{Introduction}

Young Stellar Objects (YSOs) represent a crucial stage in stellar evolution, encompassing protostars still embedded in their natal clouds to pre-main sequence stars with \ACG{still} active accretion disks. \ACG{Accretion o}utbursts in YSOs are episodic events characterized by \ACG{a} significant increase in luminosity, accompanied by enhanced mass accretion rates and outflows. These phenomena play a vital role in the mass assembly of young stars and the evolution of their circumstellar environments.

YSO outbursts are generally classified into two main categories: FU Orionis (FUor) and EX Lupi (EXor) type events. 
FUor outbursts represent some of the most dramatic variability events observed in YSOs. These outbursts are characterized by a sudden increase in brightness of up to 5-6 magnitudes in optical wavelengths, followed by a slow decay that can last for decades \citep{1977Herbig, 1996HartmannKenyon}. The prototype of this class, FU Orionis itself, underwent such an outburst in 1936 and has been slowly fading ever since (Herbig, 1966). 
EXor outbursts, named after their prototype EX Lupi, represent another class of eruptive phenomena observed in YSOs. These events are characterized by shorter-duration and more frequent outbursts compared to their FU Orionis counterparts, typically lasting from a few months to a year \citep{Herbig89, 2008Herbig}. EXors are generally associated with classical T Tauri stars, which are low-mass pre-main sequence objects that have largely emerged from their natal envelopes but are still actively accreting from their circumstellar disks.
However, recent observations have revealed a continuum of outburst behaviors, suggesting that these classifications may represent extremes of a more diverse phenomenon \citep{2014AudardAbraham, 2023Fischer}.

The primary driver of YSO outbursts is thought to be a sudden increase in the mass accretion rate from the circumstellar disk onto the central star. During quiescence, typical accretion rates for T Tauri stars are on the order of $10^{-9}$ to $10^{-7} \msunyr$ \citep{1998Gullbring}, but during outbursts these rates can increase by several orders of magnitude, reaching up to $10^{-4} \msunyr$ in extreme FUor events \citep{2014AudardAbraham}.

Several mechanisms have been proposed to explain the triggering of YSO outbursts. One possibility is thermal instability, where the disk undergoes a transition from a cool, neutral state to a hot, ionized state, leading to increased viscosity and enhanced accretion \citep{1994BellLin}. Another mechanism involves gravitational instability, which can result in the formation of dense clumps in the outer disk that migrate inward and trigger episodic accretion events \citep[e.g.,][]{2010VorobyovBasu, 2020OlivaKuiper}. A third proposed mechanism is the activation of the magnetorotational instability (MRI) in the inner disk, potentially coupled with gravitational instability in the outer regions \citep{2009ZhuHartmann}. Tidal interactions, caused by perturbations from binary companions or massive planets in the disk, have also been suggested as a trigger for outbursts \citep{1992Bonnell, NayakshinLodato12}. Finally, magnetospheric instabilities—characterized by cycles of accumulation and rapid accretion of material due to interactions between the stellar magnetosphere and the inner disk—provide another plausible explanation \citep{2010DAngeloSpruit}.

The impact of FUor outbursts on stellar evolution is significant, as these events can alter a star’s final mass, affect the structure and chemistry of the protoplanetary disk, and influence planet formation processes \citep{2018Cieza}. Recent high-resolution imaging and spectroscopic studies have advanced our understanding of FUor and FUor-like objects by revealing detailed disk structures and the dynamics of accretion processes \citep{2016LiuTakami, 2020Perez}. Moreover, the detection of FUor outbursts in both very young Class I objects and more evolved Class II sources indicates that these dramatic events can occur at various stages of \ACG{the YSO} evolution \citep[e.g.][]{2018ConnelleyReipurth}.


While outbursts have been well-documented in low-mass YSOs for decades, their occurrence in high-mass YSOs (HMYSOs) has only recently been confirmed. High-mass stars, typically defined as those with masses exceeding 8 \ACG{$\msun$}, have traditionally been thought to form through steady accretion or competitive accretion in clusters \citep{2007ZinneckerYorke}. However, recent observations have revealed that HMYSOs can also undergo episodic accretion events, challenging our understanding of massive star formation.

The first clear detection of an outburst in a HMYSO was reported by \citet{2017Caratti}, who observed a 4-magnitude brightness increase in the \ACG{near-}infrared for the 20 solar mass protostar S255IR NIRS 3.
Subsequent observations have further supported this notion, with several other cases of outbursts in massive YSOs being reported in recent years. Notable examples include the luminous outburst in NGC 6334I-MM1 detected through millimeter and centimeter wavelength observations \citep{2017Hunter}, the flaring event in G358.93-0.03 accompanied by multiple maser transitions \citep{2021Stecklum}, and the outburst in G323.46-0.08 observed in infrared and radio wavelengths \citep{2019Proven-Adzri, 2024WolfStecklum}. These events have provided compelling evidence that episodic accretion is a phenomenon that spans the entire stellar mass spectrum.

The implications of these discoveries are far-reaching. They suggest that the formation mechanism of high-mass stars may be more similar to that of their low-mass counterparts than previously thought \citep[see review by][]{2025Beuther}, involving episodic accretion from a circumstellar disk \citep{2019MeyerVorobyovElbakyan, 2020OlivaKuiper}. 
Furthermore, the presence of such outbursts implies the existence of massive, potentially unstable disks around high-mass protostars \RK{— despite earlier theoretical predictions that strong radiative feedback and photoevaporation would rapidly disperse these disks \citep[e.g.,][]{1994Hollenbach, 2002YorkeSonnhalter}. Recent high-resolution observations have, however, confirmed that massive disks can survive in these environments and become unstable \citep{2019Ahmadi, 2023Ahmadi}}.

However, the study of outbursts in HMYSOs presents unique challenges. The rarity of massive stars, their typically short formation timescales, and the high levels of obscuration in their natal environments make these events difficult to detect and study \citep{2018Motte, 2007Cesaroni}. Despite these obstacles, recent advancements in observational techniques, including long-term multi-wavelength monitoring campaigns and high-resolution interferometry, have improved our ability to detect and characterize these elusive phenomena \citep{2018Ilee, 2023ContrerasPena}.


Several mechanisms have been proposed to explain outbursts in HMYSOs. For example, disk fragmentation suggests that gravitational instabilities in massive disks can lead to the formation of clumps that rapidly migrate inward, triggering accretion bursts \citep{2017MeyerVorobyov, 2020OlivaKuiper}. Thermal instabilities, analogous to those thought to drive FUor outbursts in low-mass stars, might operate at even higher temperatures and accretion rates in high-mass systems \citep{1994BellLin}. Additionally, magnetic interactions — though their role in high-mass star formation remains less clear \citep{2016Wade} — could potentially trigger episodic accretion through magnetospheric instabilities \citep{2007Pudritz}, while external perturbations in dense clusters, such as close encounters with other stars, may induce disk instabilities and trigger outbursts \citep{2006BonnellBate}.

Among the proposed mechanisms, thermal instabilities are particularly intriguing in the context of high-mass stars, whose more massive and hotter disks may readily satisfy the conditions for triggering classical thermal instability (TI) outbursts. While many previous studies have focused on low-mass T Tauri systems — examining scenarios where TI outbursts are initiated by the activation of the magnetorotational instability (MRI) in the dead zone \citep{2009ZhuHartmann, 2020Kadam, 2020VorobyovKhaibrakhmanov, 2021Steiner, 2024CecilFlock} — our work adopts a different perspective. We specifically investigate the classical TI driven by hydrogen ionization in HMYSOs \citep{1994BellLin, 2009ZhuHartmannGammie, 2024Nayakshin, 2024Jordan}. Unlike our earlier studies \citep{2021ElbakyanNayakshin, 2023ElbakyanNayakshin, 2024Nayakshin, 2024Elbakyan}, which employed a one-dimensional approach, we now implement a two-dimensional model. This advanced approach allows us to examine in detail how TI-induced accretion events impact the disk’s vertical and radial structure, offering a complementary view to earlier 1D studies and MRI-focused studies. In doing so, our work highlights how classical TI outbursts can dramatically reshape the disk structure in HMYSOs, providing new insights into the complex processes governing massive star formation.

Our objective is to determine the fundamental properties of these outbursts, including the mass accretion rate during the event, the time taken for the outburst to reach its peak, and its overall duration. Additionally, we investigate how various parameters influence the detectability of these bursts, such as the stellar mass, the viscosity, and the rate at which material is supplied to the system.

This paper is organized as follows. In Sect.~\ref{sec:model}, we present our numerical model and detail the initial conditions that form the basis of our simulations. In Sect.~\ref{sec:TI}, we focus on a detailed study of our fiducial model of TI bursts in a HMYSO disk. In Sect.~\ref{sec:observ}, we explore the observational features of TI bursts by presenting synthetic observations and comparing our results with existing observational data. Sect.~\ref{sec:discussion} discusses the broader implications of our findings, including the role of convective motion, a comparison with previous one-dimensional models, the shortcomings of our current approach, and potential alternative mechanisms that may contribute to episodic accretion. Finally, in Sect.~\ref{sec:conclusion}, we summarize our main conclusions and outline directions for future research. Sect.~\ref{sec:Parameter_space} is dedicated to a comprehensive parameter space study, where we examine the influence of various parameters on the outburst properties: \RK{the mass deposition rate and deposition radius (Sect.~\ref{sec:mdep}),} the inner disk radius (Sect.~\ref{sec:rin}), the viscous $\alpha$ parameter or turbulent viscosity (Sect.~\ref{sec:alpha}), grid resolution (Sect.~\ref{sec:resolution}), and stellar mass (Sect.~\ref{sec:st_mass}).


\section{Numerical model}\label{sec:model}

We perform 2D radiation-hydrodynamics calculations of a protostellar disk using the hydrodynamics code PLUTO \citep{2007Mignone}. The fundamental hydrodynamic equations solved describing the conservation of mass, momentum, and energy are given as 
\begin{equation}
\label{eq:cont}
    \frac{\partial\rho}{\partial t} + \nabla \cdot (\rho \boldsymbol{u}) = 0,
\end{equation}
\begin{equation}
\label{eq:mom}
    \frac{\partial(\rho \boldsymbol{u})}{\partial t} + \nabla \cdot (\rho \boldsymbol{u} \otimes \boldsymbol{u}) = -\nabla P + \rho \boldsymbol{g},
\end{equation}
\begin{equation}
\label{eq:energ}
    \frac{\partial E}{\partial t} + \nabla \cdot ((E + P) \boldsymbol{u}) =  \rho \boldsymbol{u} \cdot \boldsymbol{g},
\end{equation}
where $\rho$ is the gas density, $u$ is the gas velocity, $P$ is the thermal pressure, and $E$ is the total (internal + kinetic) energy. The acceleration term $g$ takes into account the gravitational acceleration from the star ($GM_*/r^2$), the self-gravity of the gaseous disk, found by solving Poisson's equation \citep{2010Kuiper_b, 2011Kuiper} \RK{(but only enabled in model~\textit{TI\_5} with its massive disk)}, and the shear viscosity term $\nabla\Pi/\rho$, where $\Pi$ is the viscous stress tensor implemented with the $\alpha$-parametrization of \citet{1973ShakuraSunyaev}. No bulk viscosity is considered. 
We parametrise the kinematic viscosity in the disk using the dimensionless $\alpha$ parameter as $\nu=\alpha c_{\rm{s}} H$, where $H$ is the vertical 
\RK{pressure} scale height of the disk. A wide range of values for $\alpha$ is used for astrophysical disks, from $10^{-4}$ to $\sim0.2$ \citep{2012MuldersDominik, 2016Pinte, 2017Lodato, 2017Rafikov, 2018AnsdellWilliams, 2018NajitaBergin, 2018Dullemond, 2020Flaherty, 2020Rosotti, 2023Rosotti}.

Gas thermodynamics is considered under the approximation of local thermodynamic equilibrium (LTE) and an equilibrium temperature between the gas and the radiation. Radiation transport is treated with the gray flux-limited diffusion (FLD) approximation \citep{2010Kuiper_a, 2020Kuiper}. \REF{We do not include direct stellar irradiation in the model (see Sect.~\ref{sec:limitations}).}
Following \citet{2015Vaidya}, we use a non-perfect ideal equation of state (EoS) of \citet{2013DAngeloBodenheimer} that accounts for hydrogen ionization and dissociation, using a non-constant adiabatic index. 
We use tabulated dust opacities from \citet{2003SemenovHenning} and gas opacities from \citet{2014Malygin}.
The hydrodynamic equations are solved using a shock-capturing Riemann solver within a conservative finite volume framework, while the flux-limited diffusion (FLD) equation is solved implicitly using \RK{a} generalized minimal residual solver (GMRES). We employ the Harten-Lax-Van Leer (HLLC) approximate Riemann solver, \RK{including the contact discontinuity} along with a minmod flux limiter. Time integration is carried out using the second-order Runge-Kutta (RK2) method.

\subsection{Initial conditions}
Equations (\ref{eq:cont}) -- (\ref{eq:energ}) are solved on a polar grid ($r, \theta$) with axial symmetry \RK{around $\theta=0$} and equatorial symmetry \RK{at $\theta=\pi/2$}. We use logarithmically spaced cells in \RK{the} radial direction and either cosinusoidally or uniformly spaced cells in \RK{the} polar direction. The cosinusoidal spacing provides higher resolution near the disk midplane. The computational domain consists of 125 cells in the radial direction, spanning from 0.08 au to 10 au, and 45 cells in the polar direction, covering 0 to $\pi/2$ radians. The smallest cell size in the radial direction is 0.03 au.
The simulation begins with a pre-existing protoplanetary disk of 0.1~$\msun$ \ACG{confined within the inner 10~au} around a 10~$\msun$ protostar \ACG{(note that the full disk mass is expected to be much higher)}. The initial profile of gas density of the disk has the following form

\begin{equation}
    \rho(r,\theta) = \rho_0 \left(\frac{r}{r_0}\right)^p \mathrm{exp}\left(-0.5\left(\frac{\mathrm{cos}(\theta)}{h}\right)^2\right),
\end{equation}
where $\rho_0$ is the gas density at the radial distance $r_0=1$~au, $h=0.05$ is the aspect ratio of the disk, and $p=-1.5$ is the exponent of the power-law density distribution.
We assume that the disk rotates with a sub-Keplerian velocity of 
\begin{equation}
    v_{\phi}(r,\theta) = f v_{\mathrm{Kep}},
\end{equation}
where $v_{\mathrm{Kep}}$ is the pure Keplerian velocity and factor $f$ accounts for the gas pressure gradient in the disk and is defined as
\begin{equation}
    f = \sqrt{1 + (p + q) \frac{h^2}{\gamma}},
\end{equation}
with $\gamma=5/3$ adiabatic index and $q=-1$ exponent of the power-law temperature distribution.
The temperature and pressure profiles of the disk are defined as 
\begin{equation}
    T(r,\theta) = c_{\mathrm{s}}^2(r,\theta) \mu / (R_{\mathrm{gas}}\gamma),
\end{equation}
\begin{equation}
    P(r,\theta) = \rho(r,\theta) c_{\mathrm{s}}^2(r,\theta) \gamma,
\end{equation}
where $c_{\mathrm{s}} = h v_{\phi}(r,\theta)$ is the local sound speed, $\mu=2.353$ is the \DW{mean} molecular weight of the gas, and $R_{\mathrm{gas}}$ is the universal gas constant.
For the entire computational domain, we use \RK{a} dust-to-gas \RK{mass} ratio of 0.01.

\REF{At both the inner and outer radial boundaries we apply a semi–permeable, zero‐gradient (‘outflow‐only’) condition on all primitive variables, including the azimuthal velocity $v_\phi$. \REFF{With this choice we avoid}  
spurious angular‐momentum reflection, and allow material to exit the grid without back‐flow.} 
Consequently, all material that crosses the inner radial boundary is treated as having been accreted onto the central protostar. For the polar boundaries, we use an axisymmetric boundary condition at the pole and an equatorial symmetric boundary condition at the disk's midplane.

\subsection{Additional mass deposition and $\alpha$ parameter}

In addition, our model include\RK{s} an additional mass deposition into the disk. The mass is deposited into the disk at a constant rate, $\dot M_{\rm dep}$, at the midplane ($z=0$) of the disk, at a radial distance of $r_{\rm dep}\sim3.7$~au. The deposited mass is added as diffuse material, with an azimuthal velocity matching that of the disk at $r_{\rm dep}$. The polar and radial velocities of the added material are set to zero. 

It is important to note that, while the computational domain extends to 10 au, the external mass is deposited closer to the central protostar, rather than at the outer edge of the domain. This approach shortens the timescale for viscous transport of the additional material toward the protostar, thereby saving computational time. \ACG{Moreover, simulating the disk out to 10 au allows us to capture the dynamics of the inner disk region that are most relevant for TI, while also providing a well-defined outer boundary to ensure realistic mass replenishment.} \RK{I}n Sect.~\ref{sec:TI}, we compare the properties of TI bursts from models with $r_{\rm dep}\sim3.7$~au and $r_{\rm dep}=10$~au, and demonstrate that the results are similar.

\RK{Previous studies of classical TI have successfully reproduced key features of accretion bursts in dwarf novae and low-mass young stellar objects— namely, the rapid rise and slower decay of luminosity — by invoking the strong opacity jump at hydrogen ionization \citep[e.g.,][]{1994BellLin, 2004LodatoClarke}. More recent one-dimensional models have extended this framework to high-mass young stellar objects, matching the longer-duration, moderate-amplitude outbursts seen in some systems \citep{2021ElbakyanNayakshin, 2024Nayakshin}. However, these models} require not only a substantial change in opacity but also different values of the $\alpha$-parameter for the cold and hot branches of the S-curve. For that reason, following \citet{1998Hameury}, in our TI models, we define the $\alpha$ parameter as 
\begin{equation}
    \ln\alpha = \ln\alpha_{\rm cold}\, + \,\frac{\ln\alpha_{\rm hot}-\ln\alpha_{\rm cold}}{1 + (T_{\rm cr}/T)^8}
    \label{alpha_vs_T}
\end{equation}
where $T_{\rm cr}=25000$~K represents the critical temperature at which the transition between $\alpha_{\rm cold}$ and $\alpha_{\rm hot}$ occurs. These $\alpha$ parameters correspond to the cold and hot branches of the S-curve, respectively. Recent observations \citep{2012Cannizzo} and MHD simulations \citep{2014Hirose, 2018Coleman, 2018Scepi} suggest $\alpha_{\rm cold}\approx0.01-0.05$. Most global disk simulations, which typically use a magnetic field configuration similar to a zero net-flux shearing-box model, produce $\alpha_{\rm hot}$ values around 0.01-0.02 \citep{2000MillerStone, 2006Hirose, 2006FromangNelson, 2010Shi, 2010Davis, 2011Flock, 2011Beckwith, 2012Simon}. However, observations of dwarf novae and X-ray binaries estimate $\alpha_{\rm hot}$ to be closer to 0.1 \citep{1999Smak, 2001Dubus}. This discrepancy raises questions about the accuracy of using zero mean field models for simulating accretion discs \citep[see discussions in][]{KingEtal07, 2009ZhuHartmann, BaiStone13}. Recent 3D MHD models \citep{2014Hirose, 2018Coleman, 2018Scepi} indicate that the combined effects of convection and magnetorotational instability can increase $\alpha_{\rm hot}$ to values consistent with those observed, being of the order of $\approx0.1-0.2$. These values were found to work relatively well for both observations \citep[e.g.,][]{2001Lasota, 2012KotkoLasota} and first-principle simulations of TI in discs \citep[e.g.,][]{2014Hirose}.

\section{Analysis of TI outbursts}\label{sec:TI}

One potential mechanism responsible for episodic accretion outbursts is the well-known phenomenon of disk thermal instability \citep[e.g.,][]{1994BellLin}, which arises from a sudden increase in opacity as hydrogen ionizes at temperatures around $\sim10^4$~K. In this scenario, the inner, thermally unstable region of the disk periodically oscillates between two stable states—the cold and hot branches of the S-curve \citep[see Sect.~6 in][]{2015Armitage}—leading to significant variability in the accretion rate onto the star \citep[e.g.,][]{2004LodatoClarke,2021ElbakyanNayakshin}. This phenomenon was initially explored within the context of dwarf nova models \citep{1979Hoshi, 1981Meyer, 1985Lin}.

\begin{table}
\center
\caption{\label{tab:1} Main properties of the models.}

\begin{tabular}{cc}
\hline 
\hline 
Model & Parameters \tabularnewline
\hline 
\multirow{6}{*}{\parbox{1.5cm}{\centering TI\_1 \\ (fiducial)}} & $M_*=10\msun$\tabularnewline 
                       & $M_{\rm disk}\RK{(r<10~\rm{au})}=0.1\msun$ \tabularnewline 
                       & $\alpha_{\rm c}$/$\alpha_{\rm h} = 0.05/0.5$ \tabularnewline
                       & $r_{\rm in}=0.08$~au \tabularnewline
                       & $\dot{M}_{\rm dep}=8\times10^{-4} \msunyr$ \tabularnewline
                       & \RK{$r_{\rm dep}=3.7$~au} \tabularnewline
\hline
 & \tabularnewline
 & \tabularnewline
\hline 
\hline 
Model & Diff\RK{erence} from fiducial \tabularnewline
\hline 
TI\_2 & $r_{\rm dep}=10$~au\tabularnewline
TI\_3 & $\dot{M}_{\rm dep}=3\times10^{-4} \msunyr$ \tabularnewline
TI\_4 & $\dot{M}_{\rm dep}=5\times10^{-3} \msunyr$ \tabularnewline
TI\_5 & $M_{\rm disk}=2.0\msun$; $\dot{M}_{\rm dep}=3\times10^{-4} \msunyr$ \tabularnewline
TI\_6 & $r_{\rm in}=0.04$~au \tabularnewline
TI\_7 & $r_{\rm in}=0.12$~au \tabularnewline
TI\_8 & $r_{\rm in}=0.16$~au \tabularnewline
TI\_9 & $\alpha_{\rm h}$ = 0.1 \tabularnewline
TI\_10 & $\alpha_{\rm c}$ = 0.2 \tabularnewline
TI\_11 & x2 resolution \tabularnewline
TI\_12 & x4 resolution \tabularnewline
TI\_13 & x8 resolution \tabularnewline
TI\_14 & $M_*=5\msun$; $r_{\rm in}=0.04$~au \tabularnewline
TI\_15 & $M_*=20\msun$; $r_{\rm in}=0.12$~au \tabularnewline
\hline 
\end{tabular}
\end{table}

\begin{figure}
    \centering
    \includegraphics[width=1\columnwidth]{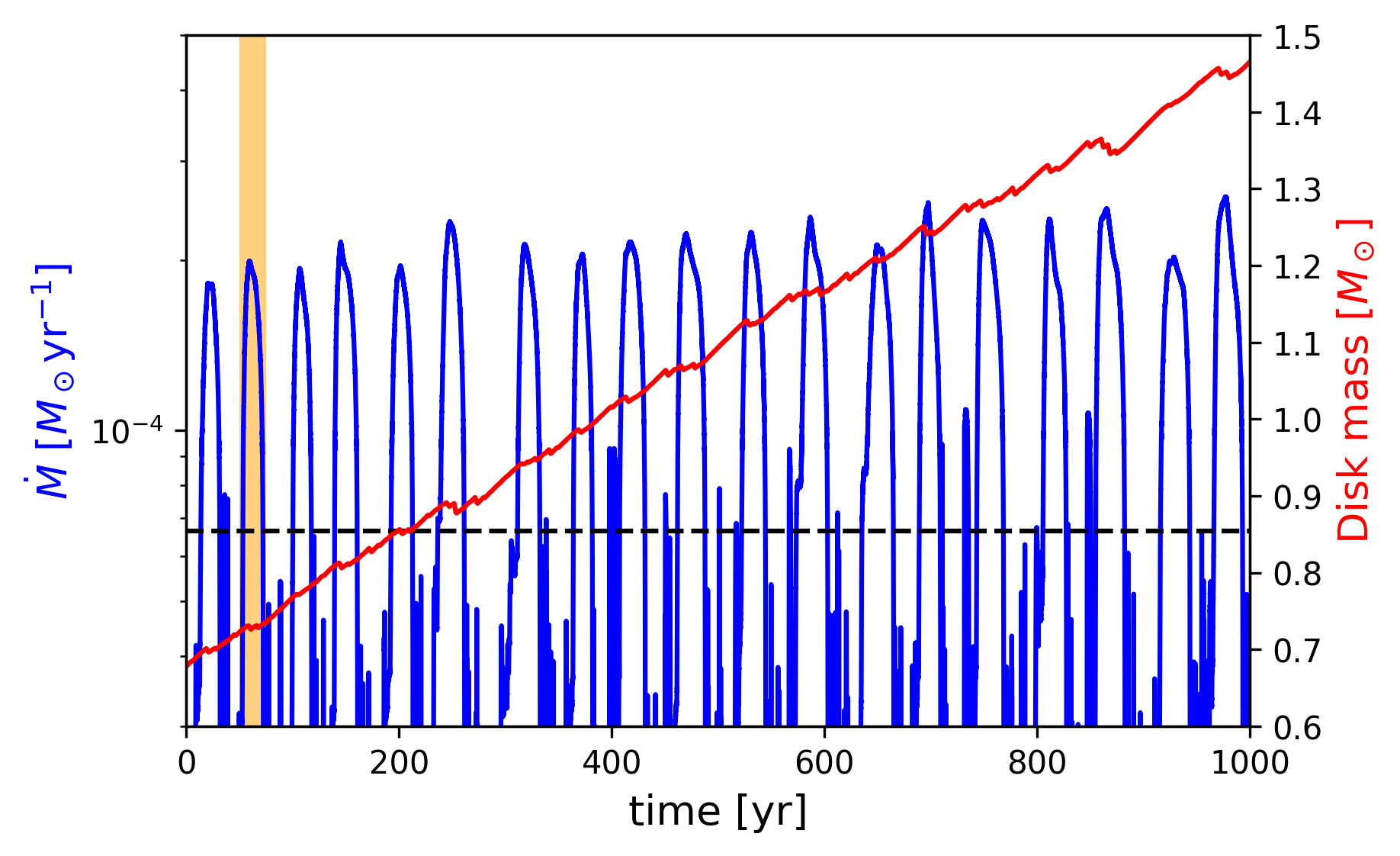}
    \includegraphics[width=1\columnwidth]{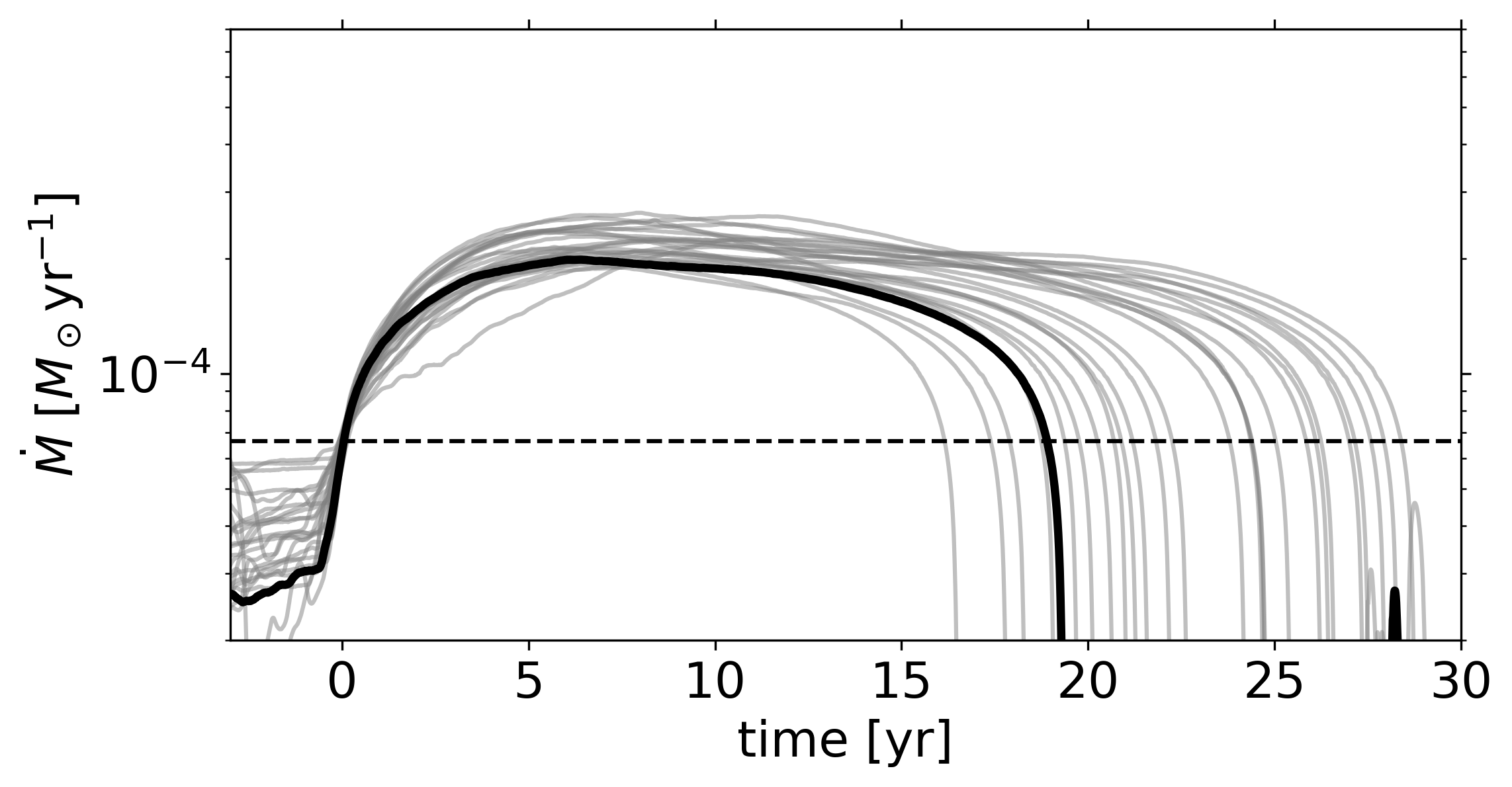}
    \caption{Mass accretion history and TI outbursts in our fiducial model TI\_1. \textbf{Top panel:} The blue line (left axis) shows the mass accretion rate, while the red line (right axis) displays the protostellar disk mass over 1000 years of stellar evolution, revealing the periodic occurrence of TI outbursts. The time origin, $t=0$, is chosen arbitrarily. \textbf{Bottom panel:} A detailed view of a single TI outburst (highlighted in orange in the top panel) is shown with a thick black line, along with subsequent outbursts in the same model (gray lines). Here, $t=0$ is defined as the moment when the accretion rate exceeds the threshold $\dot{M}_{\rm th}$ (indicated by the horizontal dashed line in both panels).}
    \label{fig:mdot_all}
\end{figure}

In the top panel of Figure~\ref{fig:mdot_all} we show 1000 years long evolution of mass accretion rate onto the HMYSO for our fiducial model (\textit{TI\_1}) with $M_* = 10 \, \msun$, the value of $\alpha_{\rm cold} = 0.05$ and  $\alpha_{\rm hot} = 0.5$, and $\dot M_{\rm dep} = 8\times10^{-4}~\msunyr$, which falls within the observed range of gas infall rates in high-mass star-forming regions. These rates are estimated to vary between $10^{-5}$ and $10^{-3}$~$\msunyr$ \citep{2013Beuther, 2017Beuther, 2021Moscadelli, 2024WellsBeuther, 2024ZhangCyganowski}. In our simulations, we adopt values of $\dot M_{\rm dep}$ and the viscosity parameters at the upper end of their observed and theoretical ranges. This choice allows us to test the upper limits of thermal instability and assess its maximum capability to generate outbursts. Since lower values of these parameters would inevitably produce less intense or shorter-duration bursts, exploring the extreme case provides a useful benchmark for evaluating the potential of TI-driven accretion variability. The time moment $t=0$ in the figure is chosen arbitrarily. The TI models with various initial properties are listed in Table~\ref{tab:1} and are discussed in the next section. 

The TI bursts in \textit{TI\_1} model have peak mass accretion rates onto the HMYSO, $\dot{M}_{\rm peak}$, of about $2-3\times10^{-4}~\msunyr$. It is worth noting that $\dot{M}_{\rm peak}$ is a factor a few lower than $\dot M_{\rm dep}$ because most of the additional mass added to the disk accumulates in the disk itself. The bulk of the disk mass is located further out, and only a small, less massive inner region is involved in a TI burst. In the top panel of Figure~\ref{fig:mdot_all} with the red line we show the total mass of the disk in the model. Clearly, only a small fraction of a disk mass is accreted onto the protostar during a TI burst. 
\ACG{Although the disk mass appears to increase over time, this trend is expected given that our model does not assume a steady-state disk, but rather follows its evolving, dynamic nature.} \REF{For a detailed analysis of the mass transport rates and characteristic timescales that explain why the disk cannot maintain in steady-state, see Appendix~\ref{sec:app_disk_mass}.}

Additionally, material from the inner disk is continuously ejected, and its ejection rate increases significantly during TI bursts. During these events, not only is material rapidly accreted onto the protostar, but a substantial fraction is also expelled to larger radial distances or driven away as strong mass outflows. \ACG{Theoretical estimates suggest that the ratio of the outflow rate to the accretion rate is typically around 0.1, although this value can vary widely from 0.01 to 1 \citep[e.g.,][]{2014Frank}}. This phenomenon is discussed further in this section. 

\ACG{In practice, one can directly monitor the mass accretion rate from spectroscopic lines (e.g. Br$\gamma$ or other hydrogen lines). Nonetheless,} in our study we compare the accretion luminosity with the stellar luminosity as a proxy for detectability. Since the total bolometric luminosity is given by $L_{\rm bol} = L_* + L_{\rm acc}$ and typically $L_{\rm acc} > L_*$ during an outburst, a TI burst is considered observable when the increase in accretion causes $L_{\rm acc}$ to become comparable to or exceed $L_*$. For simplicity, we introduce a threshold mass accretion rate, $\dot{M}_{\rm th}$, defined by
\begin{equation}
    \dot{M}_{\rm th}=\frac{L_*R_*}{GM_*},
    \label{eq:mdot_obs}
\end{equation}
where $L_* = 1.4\,(M_*/\msun)^{3.5} L_\odot$ is the stellar luminosity, $R_* = (M_*/\msun)^{0.7} R_\odot$ is the stellar radius, and $M_*$ is the stellar mass. This threshold roughly corresponds to the condition $L_{\rm acc} \gtrsim L_*$.

Here, we use mass-luminosity and mass-radius relations for ZAMS stars to determine the threshold value. This value is considered a conservative minimum, as it does not account for changes in stellar luminosity and radius during intense mass accretion. During an accretion outburst, with high accretion rates of $\dot M_*>10^{-4}~\msunyr$, the radius of a HMYSO can increase dramatically, reaching several hundred solar radii \citep{2009HosokawaOmukai, 2013KuiperYorke, 2017TanakaTan, 2019MeyerVorobyov}. In contrast, during more moderate accretion with $\dot M_*<10^{-4}~\msunyr$, stars with masses $M\gtrsim10$~$\msun$ have radii close to their ZAMS value. To avoid discrepancies and the introduction of new free parameters, we set $R_*=r_{\rm in}=0.08$~au$\approx17$~$\rsun$, which represents an intermediate value between these extremes. However, when calculating $\dot{M}_{\rm th}$, we use the relation $R_*\propto M_*^{0.7}$, as it is a more conservative estimate that prevents underestimating the number or duration of TI bursts. In future studies, we plan to explore accretion outbursts on HMYSOs with the self-consistent evolution of stellar radius using a stellar evolution code. \RK{The threshold $\dot M_{\rm th}$ is shown in Figure~\ref{fig:mdot_all} with the horizontal dashed line.  Variations in $\dot M$ below this line may not produce a detectable change in the stellar luminosity alone, but because the IR output from the heated inner disk can far exceed the stellar photosphere — even modest accretion increases can still yield observable infrared brightening.}

The top panel of Figure~\ref{fig:mdot_all} demonstrates that TI bursts can occur repeatedly in HMYSOs over their formation timescale of a few $\times10^5$ years \citep{2021Sabatini}. In the bottom panel, the thick black line represents the TI outburst highlighted in orange in the top panel, which is compared with 20 subsequent bursts from the same TI\_1 model. The subsequent bursts are shifted to have the same starting point for better visual comparison. In our simulations, these TI bursts typically last between 15 and 30 years, with recurrence timescales of approximately 20 to 50 years. We determine the burst duration by measuring the time interval between when the mass accretion rate first exceeds the threshold $\dot{M}_{\rm th}$ and when it subsequently drops below that threshold. Notably, the peak accretion rate, $\dot{M}_{\rm peak}$, varies only slightly—from $1.9\times10^{-4}$ to $2.6\times10^{-4}~\msunyr$. This limited variation likely results from the bursts being triggered within a relatively narrow inner disk region, where the mass reservoir and local physical conditions remain nearly constant between events.

\begin{figure}
    \centering
    \includegraphics[width=1\columnwidth]{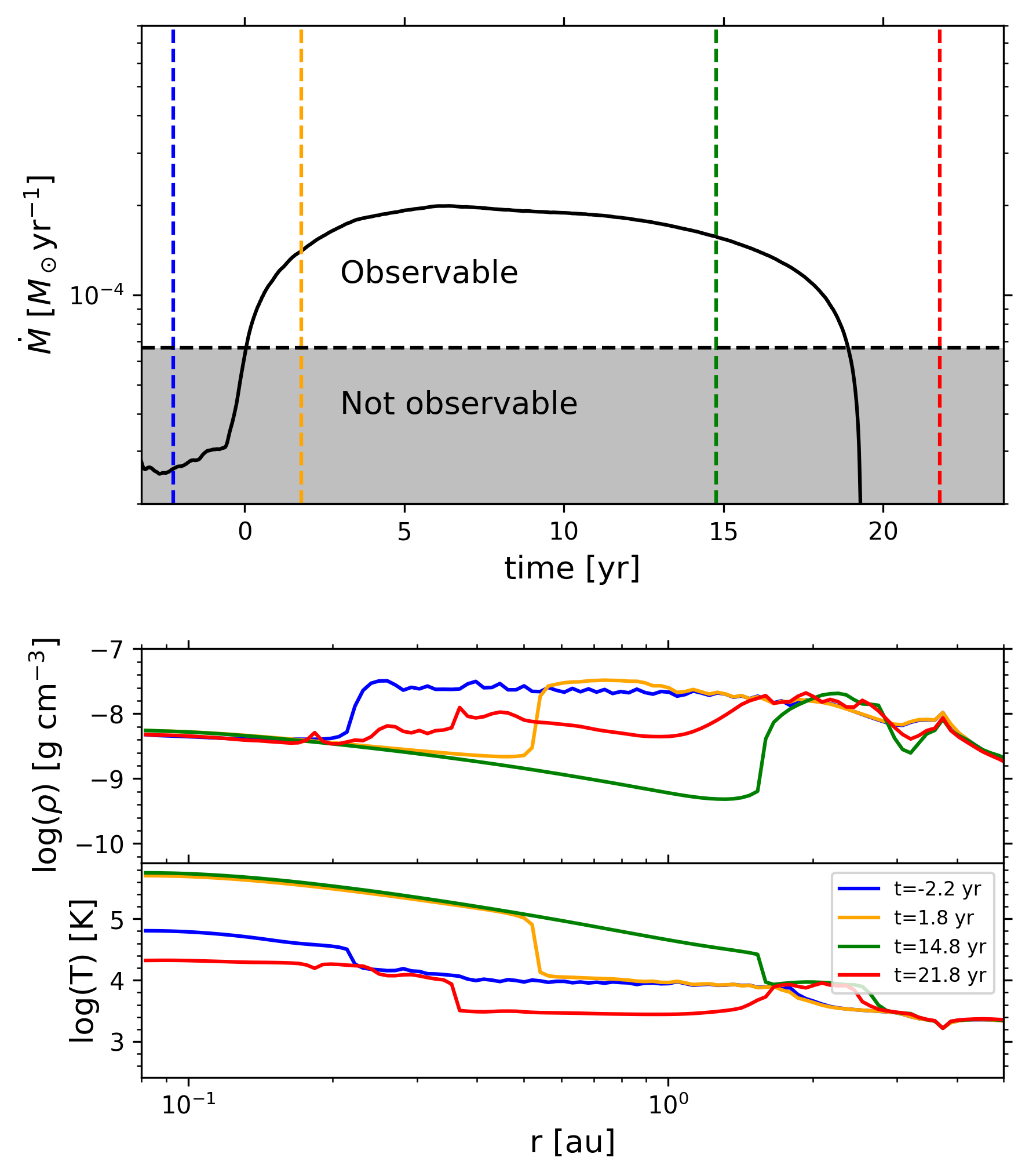}
    \includegraphics[width=1\columnwidth]{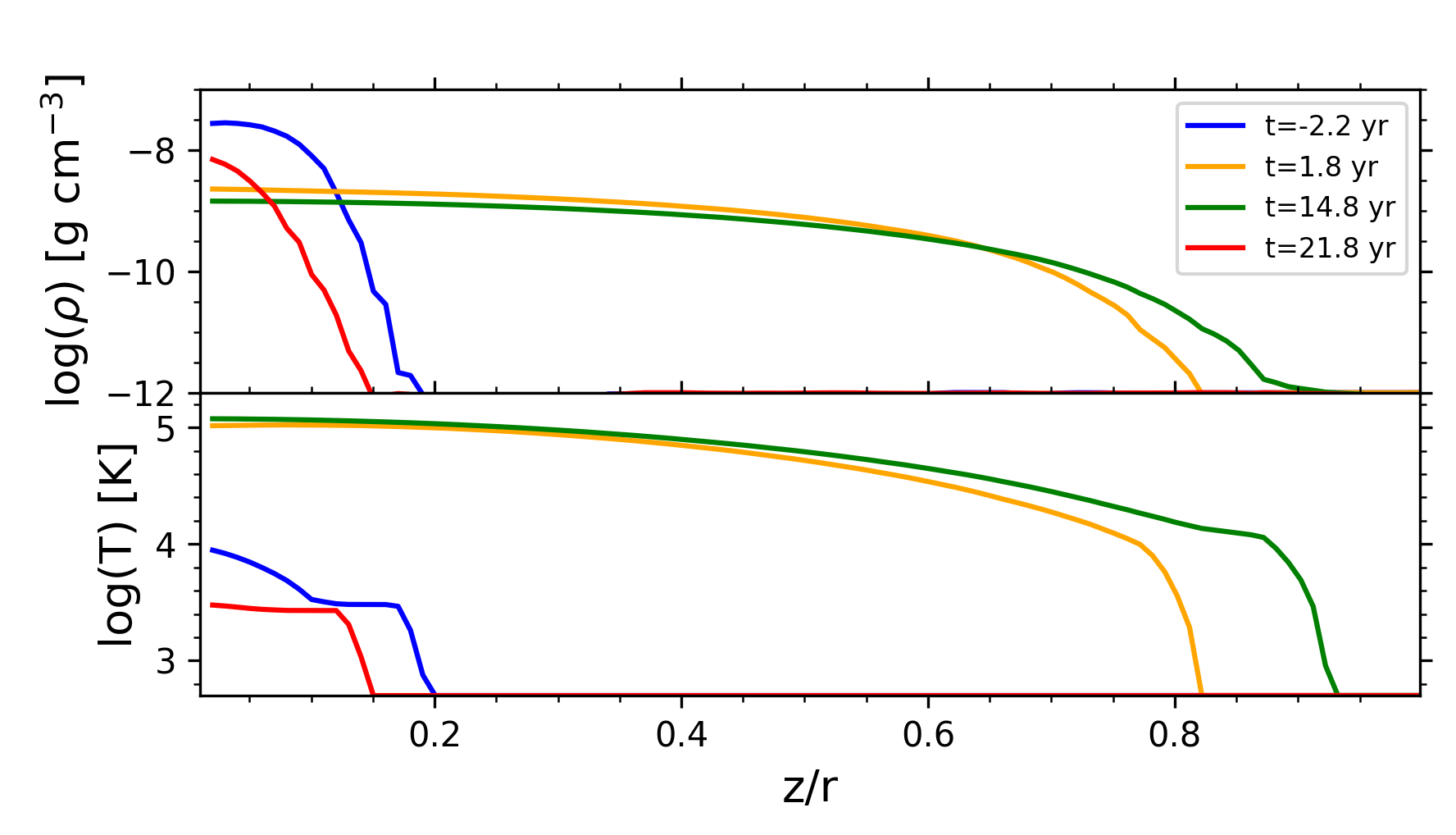}
    \caption{Mass accretion rate and spatial distribution of disk density and temperature during a TI outburst. \textbf{Top panel:} The mass accretion rate is shown for the TI outburst (highlighted in orange in Figure~\ref{fig:mdot_all}). The gray shaded region below the threshold accretion rate, $\dot{M}_{\rm th}$ (dashed horizontal line), indicates the range of variability that remains unobservable. Vertical dashed lines mark the time instances corresponding to the spatial distributions presented in the middle and bottom panels with the corresponding colors. \textbf{Middle panel:} Radial profiles of density and temperature along the disk midplane are displayed before, during, and after the burst. \textbf{Bottom panel:} Vertical profiles of density and temperature at a radial distance of $r=0.5$~au are shown for the same epochs, illustrating the disk’s vertical structure evolution.}
    \label{fig:mdot_burst}
\end{figure}

In a quasi-steady state, TI bursts are highly periodic and repeatable. Here, we focus on a single TI outburst from \textit{TI\_1} model, highlighted orange in the top panel of Figure~\ref{fig:mdot_all} to study its properties in more detail. The solid black curve in the top panel of Figure~\ref{fig:mdot_burst} shows the mass accretion rate onto the star vs time for this model. The horizontal dashed line shows the value of $\dot{M}_{\rm th}$. The variability of the accretion rate in the shaded region below that line will not be visible. The middle panel of Figure~\ref{fig:mdot_burst} presents the radial profiles of gas density and temperature in the midplane of the disk, at four distinct time moments indicated with the corresponding vertical lines in the top panel. The time moment $t=0$ in the figure corresponds to the time when the accretion rate exceeds $\dot{M}_{\rm th}$.

The TI burst begins close to the inner edge of the disk, around a radial distance of $\sim0.1$~au, where the temperature rises to the hot stable branch and hydrogen becomes ionized. The ionization front then propagates outward through the disk \citep[e.g.,][]{1994BellLin}, pushing the disk onto the hot stable branch at increasingly larger radii. However, $\dot M_*$ continues to decrease steadily over time. Eventually, the ionization front stalls at $r_{\rm TI}\approx 2$~au before retreating as the disk returns to a quiescent state. The outward propagation of the ionization front creates a region with increased density and temperature where the front stalls. This outcome is consistent with findings from \citep{2024Elbakyan}, where TI bursts were studied using a 1D approach.

At the end of the outburst, the surface density profile of the disk does not return to its pre-burst state (compare the blue and red curves \ACG{in Figure~\ref{fig:mdot_burst}}). This occurs because the disk is depleted onto the star during the burst. A subsequent outburst will only occur once the inner disk region is replenished with fresh material from the outer regions. The burst lasts about 19 years, which is consistent with the viscous timescale of the disk at the outer edge of the ionized region.

\begin{figure}
    \centering
    \includegraphics[width=1\columnwidth]{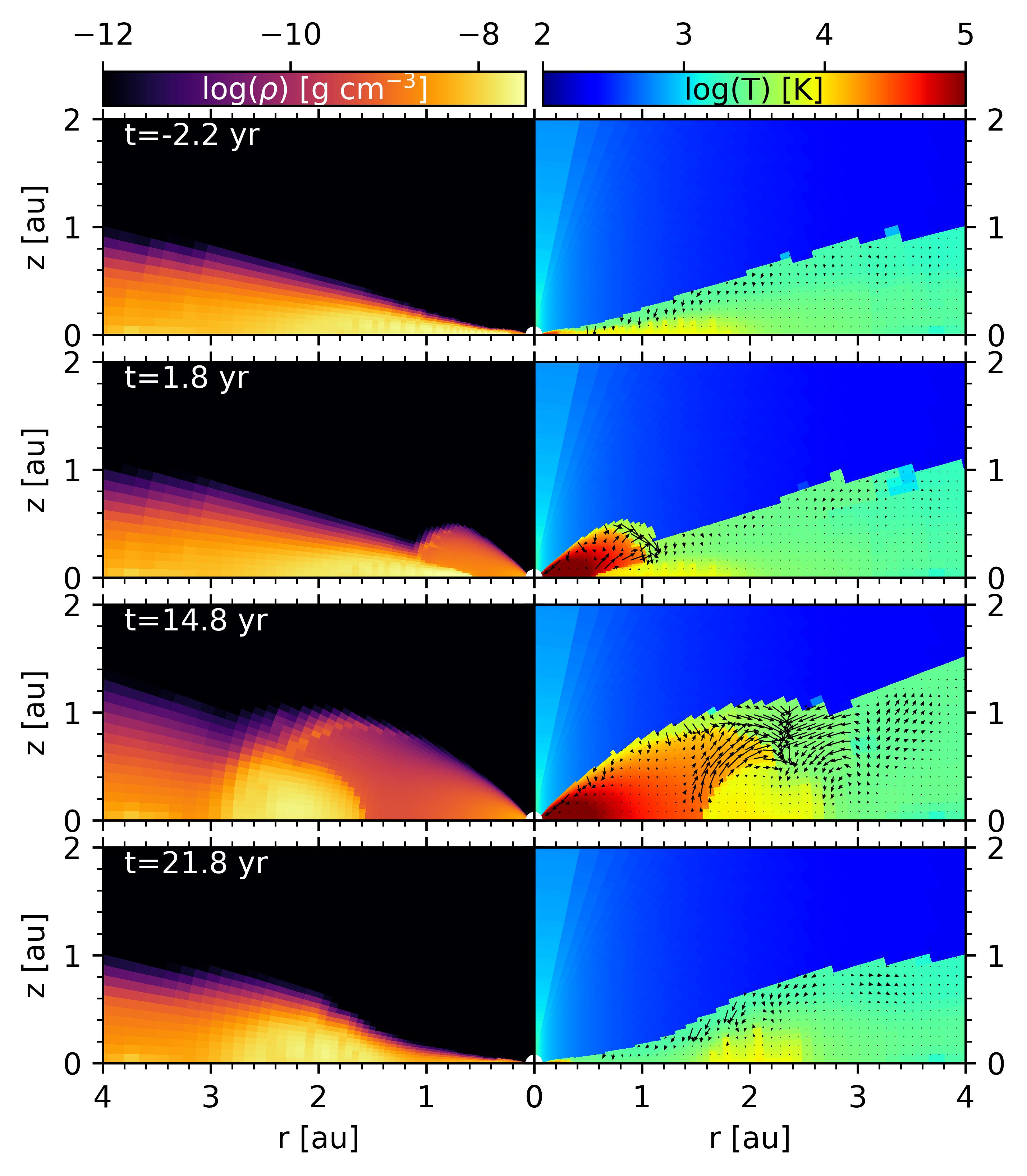}
    \caption{Density and temperature during the bursts in 2D
    Two-dimensional distributions (r-z plane) of density (left panels) and temperature (right panels) at four different time instances during a TI outburst. \RK{Black arrows in the right panels show the total velocity field.} As the outburst progresses, the inner disk heats up and expands vertically, while the density structure adjusts to the higher temperatures. Later, the disk begins to cool and partially recollapse, illustrating the transient nature of TI-driven events in HMYSO disks.}
    \label{fig:dens_temp_2d}
\end{figure}

The bottom panel in Figure~\ref{fig:mdot_burst} illustrates the vertical (z-axis) distribution of gas density and temperature at a radial distance of $r=0.5$~au during the TI burst. It is clear that during the burst, the disk undergoes significant vertical expansion, reaching a height-to-radius \RK{aspect} ratio of $z/r\sim1$. This dramatic puffing up of the disk is a direct result of the thermal instability, which causes the inner regions of the disk to become thermally unstable and expand.

To gain a deeper understanding of the vertical structure of the disk during a TI outburst, in Figure~\ref{fig:dens_temp_2d} we present a vertical (r-z) slice of gas density (left panels) and temperature (right panels) from \textit{TI\_1} model, focusing on the inner 4~au of the disk. The snapshots in the figure correspond to the same time instances for which the radial and vertical profiles are shown in Figure~\ref{fig:mdot_burst}. 
At $t = -2.2$ yr, before the outburst begins, the disk is relatively thin, with a cooler, denser inner region gradually tapering off in the radial direction. Notably, shortly after the burst begins ($t = 1.8$~yr), several key phenomena become evident. First, outflows are seen emerging from the inner regions of the disk, driven by the intense energy released during the burst. Additionally, the inner disk is observed to puff up significantly, as the increase in temperature forces the disk material to expand vertically. Lastly, the ionization front propagates outward, pushing the disk material further from the star, contributing to the redistribution of mass within the disk during the outburst. During the outburst, around $t = 14.8$~yr, the temperature distribution reveals a “puffed-up” inner disk, while the density shows a notable contrast between the heated inner region and the cooler outer disk. 

\begin{figure}
    \centering
    \includegraphics[width=1\columnwidth]{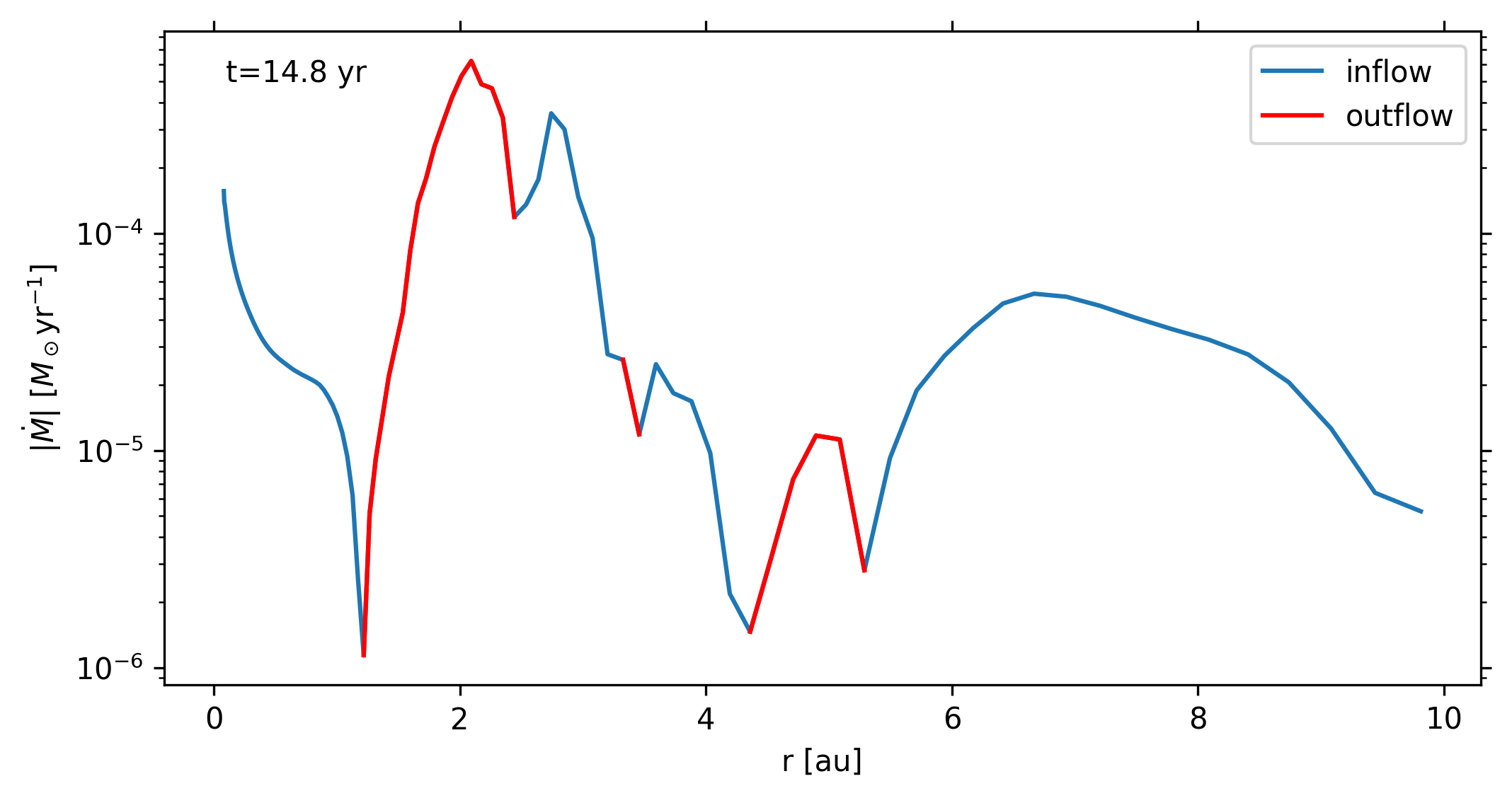}
    \caption{\ACG{Absolute mass transport rate as a function of radial distance at $t = 14.8$~yr, integrated over the polar direction. Inflow (blue) represents material moving inward toward the star, while outflow (red) represents material being driven outward by the TI outburst.}}
    \label{fig:mdot_rad}
\end{figure}

\ACG{To illustrate the strength of these outflows during the TI outburst, Figure~\ref{fig:mdot_rad} shows the absolute mass transport rate at each radial distance for $t = 14.8$~yr, integrated over the polar direction. The blue curve indicates inward transport (inflow), whereas the red curve indicates outward transport (outflow). In the inner $\sim1.2$~au, mass flows inward toward the star at rates of $\sim10^{-6}$–$10^{-4}\,\msunyr$, gradually declining with radial distance and reflecting the density-depleted region near the star. Beyond 1.2~au, however, there is a pronounced outflowing region extending to about 2.5~au, where mass transport rates peak at $\sim6\times10^{-4}\,\msunyr$. These outflow rates exceed the accretion rate onto the star at the same time, suggesting that such powerful outflows can dominate the mass redistribution in this part of the disk. Observationally, detecting outflows could offer a valuable indicator of ongoing disk processes in the inner few au. Such high outflow rates may also influence subsequent accretion phases, as well as the overall evolution of the disk and potential planet-forming regions.}

By $t=21.8$~yr, the disk has cooled to temperatures below those at the initial stage and become less dense in the inner region. This indicates that the system does not simply revert to its original pre-burst structure; instead, the outburst leaves behind a more diffuse, partially depleted inner disk. \ACG{This effect is linked to a sharp drop in the mass accretion rate at the end of the outburst. The accretion rate is higher just before a burst and then drops sharply at the end before gradually recovering until the next event is triggered (e.g., see Fig.~\ref{fig:arate_multi_mass}). Although during quiescence the star dominates the emission and may mask the full extent of the decline, the observed decrease in accretion—such as the approximately 1 magnitude drop in the K-band observed in S255IR NIRS3 \citep{2023Fedriani} — supports our findings. While the observed timescales are somewhat shorter, this decline in accretion reinforces the idea that the burst can leave a lasting impact on the inner disk structure and accretion behavior. It remains an open question whether other burst mechanisms, such as gravitational instabilities, might also produce a similar drop in accretion.}

Overall, the spatial distributions of density and temperature demonstrate how a TI outburst can significantly alter both the radial and vertical disk structure in a relatively short time. They also highlight that the return to quiescence is not instantaneous: even after the burst ends, the disk continues to adjust in terms of density and temperature, underscoring the transient yet impactful nature of TI-driven events.

\section{Observational features} \label{sec:observ}
\begin{figure}
    \centering
    \includegraphics[width=1\columnwidth]{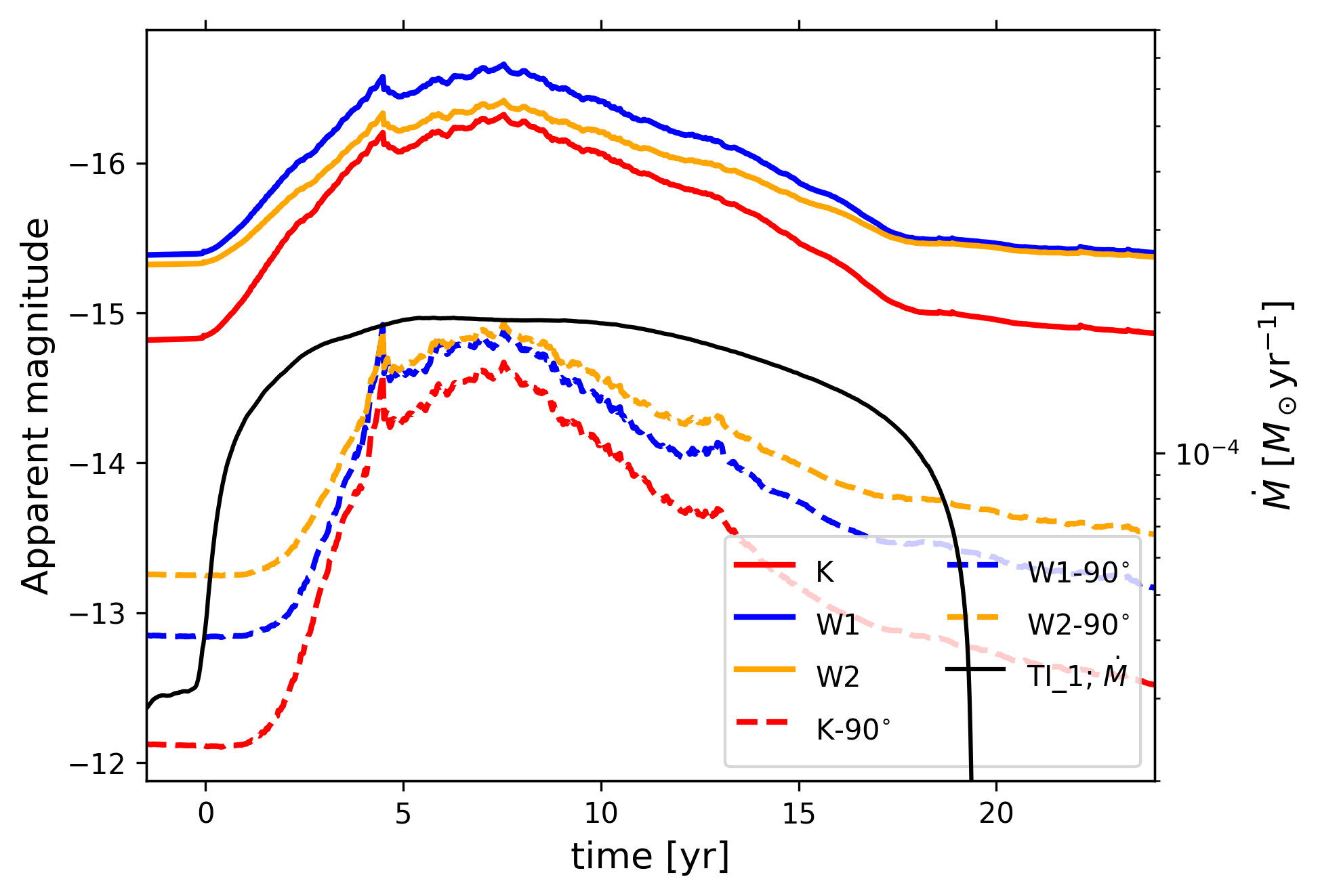}
    \caption{Infrared light curves of the TI outburst in model~\textit{TI\_1}. The curves represent the K (red), W1 (blue), and W2 (orange) bands, with the black solid line showing the accretion rate, $\dot{M}$, for reference. The solid lines correspond to a face-on disk orientation, while additional light curves (dashed lines) illustrate the effect of a 90° (edge-on) inclination. 
    }
    \label{fig:arate_obs}
\end{figure}

\subsection{Generation of synthetic observations}

To derive synthetic fluxes in various infrared bands, we perform radiative transfer calculations using RADMC-3D \citep{2012Dullemond}. The results of our two-dimensional (2D) \RK{axially symmetric radiation-}hydrodynamic simulations serves as input for RADMC-3D, which revolves the disk structure around the central axis and thus reconstruct the three-dimensional (3D) density and temperature distributions. \ACG{We note that our model does not include an envelope or outflow cavities, which can affect the observed flux by altering the overall geometry and radiative transfer. Future work will address these components to provide a more complete picture of the system.}

We adopt DSHARP dust opacities \citep[e.g.,][]{2018Birnstiel} to account for absorption, scattering, and reemission by dust grains. For simplicity, we place the system at a distance of 1 parsec. RADMC-3D computes the spectral energy distribution (SED) of the system by ray-tracing through the 3D disk. We then convolve the emergent SED with the respective filter transmission curves for the K, W1, and W2 bands to obtain synthetic fluxes in each bandpass. Repeating this procedure at different time snapshots — corresponding to various stages of the TI outburst — yields a set of time-dependent light curves that reflect how the disk’s thermal and density structures evolve under episodic accretion events.

Figure~\ref{fig:arate_obs} presents the evolution of infrared light curves in the K (red), W1 (blue), and W2 (orange) bands during a TI outburst in model~\textit{TI\_1}. The solid black line represents the accretion rate, $\dot{M}$, providing a reference for the timing and intensity of the outburst. The light curves exhibit a rapid rise phase, reaching peak brightness within approximately five years, followed by a more gradual decay lasting nearly 15 years. This timescale is consistent with the evolution of the accretion rate, which also rises sharply but begins to decline earlier than the infrared flux. The slight delay between the peak accretion rate and peak infrared emission suggests that the radiative response of the disk and envelope to the sudden increase in mass accretion is not instantaneous.  
The K-band and W1-band light curves follow each other closely, while the W2 band shows a slightly different behavior during the peak and decay phases, likely due to variations in thermal dust emission at different wavelengths, which reflect local temperature variations within the disk.

The amplitude of the outburst reaches approximately 1.5 magnitudes across all bands, indicating a significant brightening event. To assess whether this brightening will be observable, we calculate the theoretical magnitude, $m_{\rm th}$, for each band by directly computing the monochromatic luminosity using the Planck function. Specifically, for each band we evaluate the spectral radiance $B(\nu,T)$ at the band’s central frequency and then determine the luminosity per unit frequency as

\begin{equation}
    L_\nu = 4\pi^2 R_*^2\, B(\nu,T).
\end{equation}
assuming a temperature $T = 20000$~K and a stellar radius given by $R_*=(M_*/\msun)^{0.7} R_{\odot}$. The observed flux is then obtained by
\begin{equation}
    F_{\nu, \rm{obs}} = \frac{L_\nu}{4\pi d^2},
\end{equation}
where $d = 1$ pc is the distance to the source. We convert this monochromatic flux to a magnitude using
\begin{equation}
    m_{\rm th} = -2.5 \log_{10}\left(\frac{F_{\nu, \rm{obs}}}{F_0}\right),
\end{equation}
with $F_0$ being the zero-point flux for the band. \VE{Our calculations yield theoretical magnitudes of approximately -6.91, -6.86, and -6.88 for the K, W1, and W2 bands, respectively — essentially a uniform value of about -6.9.} These results confirm that the burst will be observable in all bands.

The light curves shown as solid lines in Figure~\ref{fig:arate_obs} are derived assuming a face-on disk orientation. To explore the effect of disk inclination, we performed additional radiative transfer simulations with a disk inclination of 90°. When viewed edge-on, the disk becomes highly optically thick, causing significant self-absorption and extinction. 
However, because the detectability threshold $m_{\rm th}$ is much lower, outbursts remain visible in all bands even for an edge-on configuration.  Thus, despite the strong orientation-dependent effects, the outburst remains observable across all bands.

\RK{Although the burst accretion rate is only a few times $\dot{M}_{\rm th}$, the resulting infrared magnitudes reach $\sim -16$, much higher than $-6.9$. This difference arises because $m_{\rm th}$ is based on the assumption that all accretion luminosity emerges from the stellar surface alone. In our full radiative-transfer models, however, the heated disk, spanning several astronomical units, reprocesses and emits this energy over a vastly larger area. Even a modest increase in $\dot{M}$ therefore produces a dramatic infrared brightening once the entire disk surface contributes to the emission.}

\VE{In our simulations, the burst amplitude varies between bands and depends on the viewing geometry. For the face-on case, the K-band magnitude changes by approximately 1.51 mag, the W1 band by about 1.28 mag, and the W2 band by roughly 1.10 mag. In contrast, when the disk is viewed edge-on, the burst amplitude increases: the K-band varies by approximately 2.82 mag, the W1 band by about 2.10 mag, and the W2 band by roughly 1.69 mag. The overall brightness in the edge-on configuration is significantly fainter than in the face-on case. }

\VE{While the face-on light curves in all three bands show similar behavior, the edge-on case exhibits a notable difference in the relative brightness of the bands. In the face-on orientation, the W1 band reaches the highest peak brightness (i.e., the lowest magnitude), followed by W2 and then K. However, in the edge-on configuration, the W2 band becomes brighter than the W1 band. This change likely arises from the wavelength-dependent effects of extinction in the optically thick inner disk, where the longer-wavelength W2 emission is less suppressed than the shorter-wavelength emission.}

A notable feature of the light curves is the extended plateau following the peak, \RK{reflecting} that the inner disk remains in a high-luminosity state for a prolonged period before gradually cooling and fading back to its quiescent level. This prolonged decay is consistent with the expectation that, once the inner disk heats up and transitions to the hot branch of the S-curve, it requires considerable time to cool sufficiently to return to a low state. Overall, these results demonstrate that TI-driven outbursts can produce observable brightness variations in HMYSOs, with distinct infrared signatures.

\subsection{Comparison with observational data}

Observed bursts in high-mass young stellar objects have durations ranging from a few months to 10–15 years \citep{2015Tapia, 2017Caratti, 2017Hunter, 2019Proven-Adzri, 2021Hunter, 2021Stecklum, 2022Zhang, 2024WolfStecklum}, and some objects even exhibit multiple bursts with recurrence intervals of approximately 6–30 years \citep{2021Chen, 2023Fedriani}. While the longer outburst durations in our models, which typically last between 15 and 30 years with recurrence timescales of 20–50 years, are roughly comparable to the upper end of the observed range, significant discrepancies remain. In particular, all observed bursts—whether short or long—exhibit peak accretion rates of $\dot{M}_{\rm peak}\sim\mathrm{a \, few}\times10^{-3}~\msunyr$, which is about an order of magnitude higher than the $2-3\times10^{-4}\msunyr$ predicted by our thermal instability (TI) models \ACG{(except for model \textit{TI\_8}, which, with its larger inner disk radius, produces higher accretion rates)}. Moreover, the rise times in our simulations are approximately 5–10 years, nearly an order of magnitude longer than the observed rise times of a few weeks to a year. Additionally, Figure~\ref{fig:arate_obs} shows that our typical TI outburst results in a brightening of about 1.5 magnitudes across different bands, whereas observed HMYSO outbursts show brightening of 2–3 magnitudes.

\ACG{These discrepancies suggest that the classical thermal instability, as modeled here, is unlikely to account for the full range of observed burst phenomena, particularly the shorter-duration events. One possible explanation for the lower peak accretion rates and longer rise times in our models is the inclusion of the full vertical structure. In our two-dimensional simulations, the energy released during an outburst is distributed more gradually} \SN{and widely} \ACG{throughout the disk, leading to a less impulsive increase in the accretion rate. This effect is less pronounced or absent in one-dimensional models, which may explain why TI models in 1D sometimes yield} \SN{\RK{higher} peak rates.} \ACG{Overall, our findings indicate that the accretion rate in TI outbursts is highly dependent on the physical parameters of the system and that additional processes — such as gravitational instability, disk fragmentation, or alternative mechanisms — may be required to achieve the higher rates observed in nature.}

\section{Discussion}\label{sec:discussion}

\begin{figure}
    \centering
    \includegraphics[width=1\columnwidth]{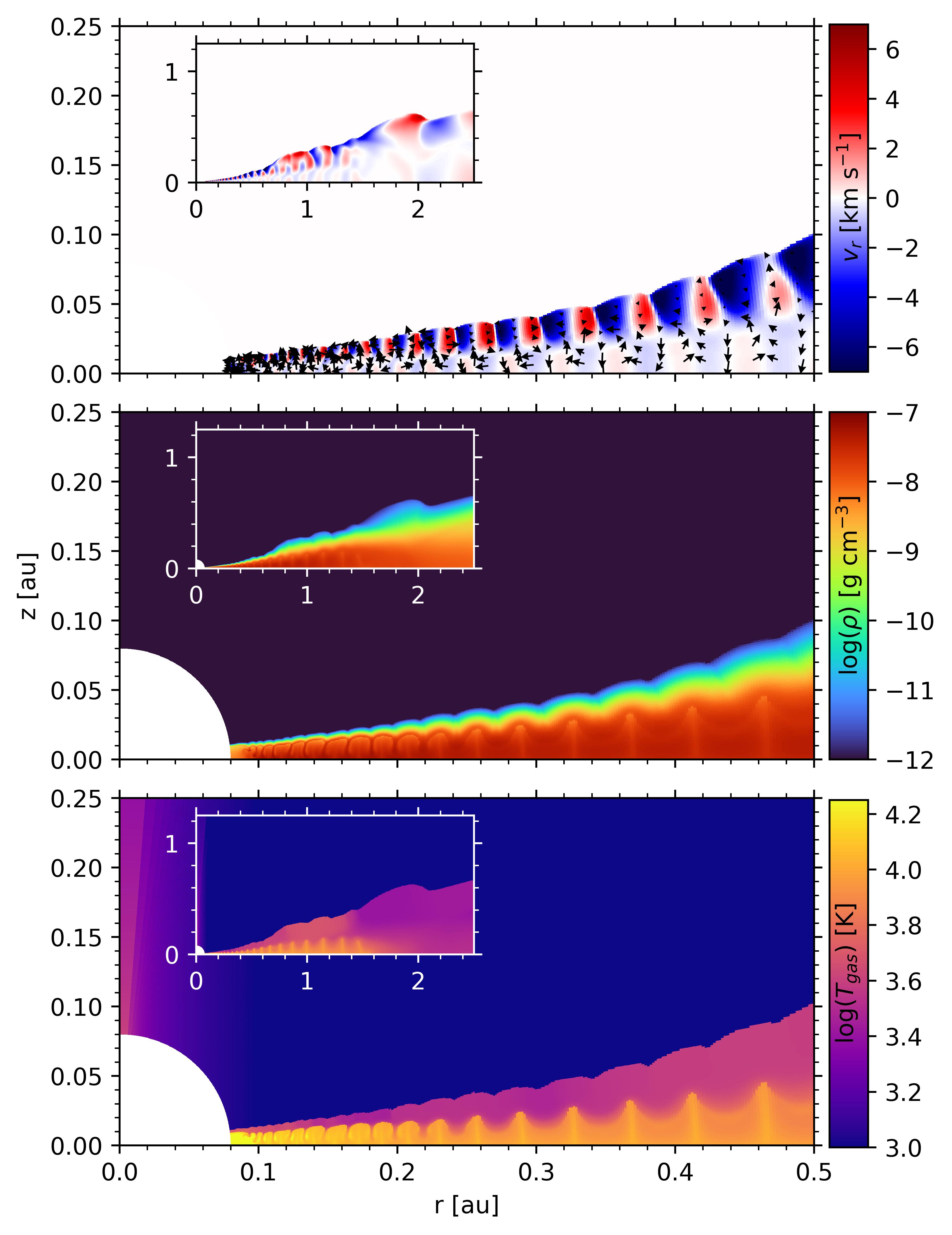}
    \caption{Two-dimensional view of the inner disk at the beginning of a TI outburst in high-resolution model TI\_13. \textbf{Top panel:} Radial velocity $v_{\rm r}$, with red indicating outflow and blue indicating inflow. Black arrows show the total velocity field. \textbf{Middle and bottom panel:} Volume density and temperature in the disk (in log scale). Insets in each panel provide a broader view of the disk for context.}
    \label{fig:dens_temp_vel}
\end{figure}

\subsection{Vertical dynamics and convective motions}

In many theoretical studies, early-stage protoplanetary disks are assumed to be in a steady state with a smooth internal structure. However, observations and recent simulations indicate that disks in their very early phases are highly dynamic and exhibit significant vertical structure \citep[e.g.,][]{2006VorobyovBasu, 2017Tsukamoto, 2017Flock, 2020Kadam, 2020OlivaKuiper, 2020Lebreuilly, 2024Speedie}. Here, we investigate in detail how the presence of vertical structure influences the triggering and evolution of TI outbursts. 

Figure~\ref{fig:dens_temp_vel} illustrates the two-dimensional structure of the disk in our highest resolution model, TI\_13, captured at the onset of a TI outburst. The top panel shows the radial velocity field with overlaid vectors representing the total velocity, while the middle panel presents the \RK{mass} volume density distribution and the bottom panel displays the \RK{gas} temperature distribution. Each panel also includes an inset that shows the same disk property over a larger spatial scale, providing context for the inner disk behavior.

A clear feature visible in the top panel is the strong convective motion within the inner $\sim$2 au of the disk. Here, viscous heating significantly raises the temperature at the midplane, while the disk surface remains relatively cool. As a result, the temperature difference between the midplane and the surface exceeds an order of magnitude in the inner disk, whereas in the outer regions ($r \gtrsim 2$~au), where viscous heating is less pronounced, the vertical temperature gradient is much shallower (less than a factor of two). This steep gradient in the inner disk drives vigorous convection, as evidenced by the eddy-like structures in the velocity vectors: \RK{hot, low-density gas rises in narrow columns from the midplane, then cools and sinks back down in the adjacent regions between these columns}. We note that similar convective motions, which give rise to these vertical structures, are also observed in our lower resolution models (see the top panels in Figure~\ref{fig:dens_temp_2d}). 
\RK{Moreover, these eddy‑like cells and hot columns are not strictly stationary: during the time they slowly drift inward along with the global accretion flow.} \SN{Previous 3D localised (shearing box) simulations of TI outbursts also found convection developing and contributing significantly to the energy transport in the disk \citep{2014Hirose, 2015Hirose, 2018Scepi}.}

These convective motions play a crucial role in redistributing energy and mass within the disk, and they highlight the importance of resolving the vertical structure when modeling TI outbursts. 
\SN{While many one-dimensional studies rely on a vertically averaged S‑curve formalism — complete with detailed thermal equilibrium curves and parameterized viscosity transitions \citep[e.g.,][]{2001Lasota} — they do not explicitly capture the full two‑dimensional interplay between radial and vertical flows. Our two‑dimensional approach, by contrast, directly resolves temperature and density gradients in both directions, revealing vigorous convection and dynamic stratification that can significantly alter burst properties.}
This is critical for developing a more accurate understanding of episodic accretion in young stellar objects \citep{1994BellLin, 2009ZhuHartmann}, and it emphasizes that the complex vertical dynamics in the inner disk must be taken into account to capture the full range of physical processes involved in TI-driven events.

\subsection{Comparison with 1D models}

Building on our previous work using the 1D hydrodynamical code DEO \citep{2024Elbakyan}, where we successfully reproduced the main characteristics of longer-duration TI bursts in HMYSOs and quantified the influence of parameters such as stellar mass, $\dot{M}_{\rm dep}$, and the viscosity prescription, we have now extended our investigation to a two-dimensional (2D) model that resolves the disk’s vertical structure. 
\SN{In our 1D calculations we follow the industry-standard approach \citep[e.g.,][]{1994BellLin, 2001Lasota}, in which the vertical disk structure during outbursts is assumed to resemble a steady-state solution: one tabulates equilibrium disk structures over a range of midplane temperatures, surface densities, and viscous heating rates (the $\alpha$-parameter) and then applies these tables to the non-equilibrium outbursting disk. This method has proven useful and is supported by local 3D shearing-box simulations \citep[e.g.,][]{2015Hirose}, but it cannot capture global, mass-transferring flows above the disk surface.} The 1D assumption tends to produce outbursts that are much shorter and brighter, as the energy release is confined to a single averaged layer without any vertical stratification. Moreover, the 1D models struggled to capture short-duration bursts with rapid rise times, as well as the occurrence of multiple sequential bursts observed in some systems. 

Our 2D simulations complement and refine these earlier findings by capturing the full vertical structure of the disk. They reveal convective motions and significant temperature gradients between the hot midplane and the cooler disk surface, as well as significant gas flows developing above the disk midplane (e.g., Fig.~\ref{fig:dens_temp_2d}). These vertical processes allow the released energy to be distributed more gradually throughout the disk, leading to outbursts that tend to have longer durations and lower peak accretion rates than those predicted by the 1D models. Importantly, both our 1D and 2D models continue to face challenges in reproducing the observed short rise times and short burst durations. \REF{A more detailed comparison of the evolution of the inner disk in our 1D and 2D simulations can be found in Appendix~\ref{sec:s_curves}.} 

\SN{While more work is necessary to explore a wider parameter space, it is notable that none of our 2D outbursts produces the so-called “reflares” seen in 1D TI calculations for dwarf novae and low-mass protostar accretion \citep[e.g.,][]{2001Lasota, 2016Coleman, 2024Nayakshin}. The absence of reflares may be due to the mass transfer above the disk midplane in our models — flows that cannot exist in vertically averaged 1D discs by design.}

Ultimately, these comparisons underscore the importance of resolving the full two-dimensional structure when modeling TI-driven events in HMYSOs. While both 1D and 2D models confirm that thermal instability is a key process in shaping episodic accretion, our 2D approach offers a more realistic depiction of the disk's behavior. This added complexity allows the disk to regulate its response to thermal instability more effectively, leading to less impulsive and more extended outburst profiles. Future studies, potentially incorporating 3D effects, will further refine our understanding and help reconcile theoretical predictions with the diverse observational characteristics of HMYSO outbursts.

\subsection{Model limitations and future improvements} \label{sec:limitations}

In our simulations, we have chosen to neglect direct stellar irradiation. Although irradiation can heat the optically thin surface layers of a disk, its effect is largely confined to these outer regions and does not significantly penetrate into the deep, optically thick midplane where most of the mass resides. The thermal structure in these dense regions is predominantly governed by viscous dissipation, with energy transport mediated by radiative diffusion and convection. Convection, in particular, is driven by the steep temperature gradients established by viscous heating; while irradiation can modify the boundary condition at the disk surface, it does not substantially alter the internal temperature profile where the convective processes are active \citep{2002YorkeSonnhalter, 2007Krumholz, 2010Kuiper_a}. 

Furthermore, in our model the radiative transport is treated using the gray flux-limited diffusion (FLD) approximation \citep{2010Kuiper_b, 2020Kuiper}. This approach captures the essential physics of radiative energy transfer in the optically thick regions without the computational expense of a full angular-dependent treatment. Given that the energy budget is dominated by viscous heating and the resulting convective transport, the direct inclusion of stellar irradiation would primarily affect only the disk’s surface layers and would not drastically modify the overall internal dynamics. This simplification is particularly justified in simulations of massive star formation, where the deep disk structure is controlled by accretion-driven processes and the effects of irradiation are limited to setting a temperature “floor” at the disk atmosphere .

While incorporating a detailed treatment of stellar irradiation could \RK{ refine the temperature and dust structure in the very surface layers, its net impact on the deep, optically thick midplane — where TI and viscous heating dominate — is minimal. Moreover, accurately modeling irradiation requires assumptions about the stellar spectrum and dust properties, each of which introduces additional uncertainties and free parameters. By omitting direct irradiation, we focus on the accretion-driven processes that control TI bursts.} Future work may extend this study by including a more comprehensive treatment of stellar irradiation.

\RK{Magnetic fields are also omitted from our current models. In HMYSOs, the strength and geometry of any large-scale field remain poorly constrained. Many massive protostars lack the outer convective zones that drive solar‐type dynamos, and direct Zeeman detections are scarce \citep[e.g.,][]{2013KochukhovSudnik, 2016Wade}. By parametrizing all angular‐momentum transport with an effective $\alpha$ viscosity, we capture the net turbulent stresses, whether hydrodynamic or magnetic, without prescribing a specific magnetohydrodynamic mechanism. Importantly, the thermal instability itself is driven by the opacity and heating physics in the dense midplane, so the first-order outburst properties are largely hydrodynamic. Future extensions of this work will include non-ideal MHD effects and field-driven winds to assess how magnetic stresses and disk–wind coupling might alter the onset or evolution of TI bursts.}

\subsection{Alternative mechanisms for episodic accretion}

While our current work focuses on TI as a primary driver of episodic accretion in HMYSOs, it is important to consider alternative mechanisms that may also contribute to the observed outburst phenomena. One promising avenue involves the role of external influences such as gravitational interactions and disk instabilities. In addition to TI, gravitational instability (GI) and disk fragmentation have been proposed as mechanisms that can trigger outbursts in HMYSOs \citep{2017MeyerVorobyov, 2020OlivaKuiper, 2021ElbakyanNayakshin, 2023ElbakyanNayakshin, 2024Elbakyan}. In this scenario, the disk becomes gravitationally unstable, leading to fragmentation into dense clumps. These clumps can then migrate inward, and upon approaching the central protostar, tidal forces may disrupt them, resulting in rapid accretion events accompanied by bursts of mass ejection.

This alternative mechanism provides a compelling explanation for the short-duration, high-intensity bursts that our TI models — both 1D and 2D — struggle to reproduce. Unlike the classical TI-driven events, which generally lead to more extended and moderately bright outbursts, the accretion of fragmented clumps can produce bright outbursts with rapid rise times. Therefore, gravitational instability and disk fragmentation may operate as complementary or even alternative mechanisms to thermal instability, particularly in systems where multiple outburst episodes or rapid bursts are observed.

\RK{In addition to direct fragmentation‐driven accretion bursts, the luminosity released when a dense clump is accreted can itself heat the surrounding disk. If this accretion luminosity irradiates regions that were previously on the cold branch of the S-curve, it can raise their effective temperature enough to push them into the thermally unstable regime. In that case, a secondary TI event could develop, characterized by a slower decay of the light curve as the heated disk relaxes back to the cold state. Observationally, this would appear as an extended tail of elevated brightness following the main fragment‐driven burst.}

Future studies that incorporate GI and fragmentation into comprehensive numerical simulations will be crucial for disentangling the relative contributions of these processes. By exploring a diverse range of physical conditions and initial parameters, we can better assess whether the episodic accretion events in HMYSOs are driven by a combination of TI and GI-related processes, or if one mechanism predominates under certain circumstances. This integrated approach promises to enhance our understanding of the complex dynamics governing mass accretion and outburst activity in massive star formation.

\section{Conclusions}\label{sec:conclusion}

We have investigated the thermal instability (TI) outbursts in high-mass young stellar objects (HMYSOs) using advanced two-dimensional \RK{axially symmetric radiation-}hydrodynamical simulations that resolve the full vertical structure of the inner disk. Our models incorporate radiative transport and viscous heating to examine how vertical stratification influences the onset and evolution of TI-driven accretion bursts. By capturing the intricate interplay between radial and vertical dynamics, these simulations reveal steep temperature gradients and robust convective motions that are absent in traditional one-dimensional, steady-state approximations.  

Our results demonstrate that the inclusion of vertical structure leads to longer outburst durations and moderated peak accretion rates, as the energy released during an outburst is distributed more gradually throughout the disk. While TI can trigger episodic accretion outbursts, the mechanism as modeled in our simulations appears too weak to fully account for the observed burst properties. Our models indicate that TI-induced bursts are characterized by lower peak accretion rates — on the order of $2-3\times10^{-4}\msunyr$ — and longer durations than many observed events, which often feature peak rates of a few $\times10^{-3}\msunyr$ and more rapid rise times.  

Despite these discrepancies, TI outbursts are predicted to be observable, with clear infrared signatures and significant brightness variations, albeit with amplitudes that fall short of the observed 2–3 magnitudes. Our 2D simulations, which resolve the vertical structure of the disk, provide important refinements over the simpler 1D models by capturing complex convective motions and steep temperature gradients that moderate the burst profile. This added realism underscores the critical role of vertical dynamics in shaping the outburst behavior.

Nevertheless, our findings suggest that additional mechanisms must be at play to reproduce the full range of HMYSO outburst phenomena. Processes such as gravitational instability, disk fragmentation, and tidal interactions may contribute to the rapid, high-intensity bursts observed in some systems. Future work incorporating these mechanisms—potentially in three-dimensional simulations—will be essential for developing a more comprehensive model of episodic accretion in massive star formation.

In summary, while thermal instability is an important component of disk physics in HMYSOs, it is likely only part of a broader ensemble of processes driving episodic accretion. Our work highlights both the strengths and limitations of TI as an explanatory mechanism and paves the way for future studies aimed at reconciling theoretical models with observational data.

\begin{acknowledgements}
      \REF{We thank the anonymous referee for their thorough comments and suggestions, which significantly improved the manuscript.} We thank André Oliva for helpful discussions. R.K. acknowledges financial support via the Heisenberg Research Grant funded by the Deutsche Forschungsgemeinschaft (DFG, German Research Foundation) under grant no.~KU 2849/9, project no.~445783058. \ACG{A.C.G. acknowledges support from PRIN-MUR 2022 20228JPA3A "The path to star and planet formation in the JWST era (PATH)" funded by NextGeneration EU and by INAF-GoG 2022 "NIR-dark Accretion Outbursts in Massive Young stellar objects (NAOMY)" and Large Grant INAF 2022 “YSOs Outflows, Disks and Accretion: towards a global framework for the evolution of planet forming systems (YODA)". Z.G. is funded by ANID, Millennium Science Initiative, AIM23-001. Z.G. is supported by the China-Chile Joint Research Fund (CCJRF No.2301) and Chinese Academy of Sciences South America Center for Astronomy (CASSACA) Key Research Project E52H540301. }
\end{acknowledgements}

%
%

\bibliographystyle{aa}
\bibliography{ref_base2}

\begin{appendix}

\section{Parameter space study}\label{sec:Parameter_space}

\begin{figure}
    \centering
    \includegraphics[width=1\columnwidth]{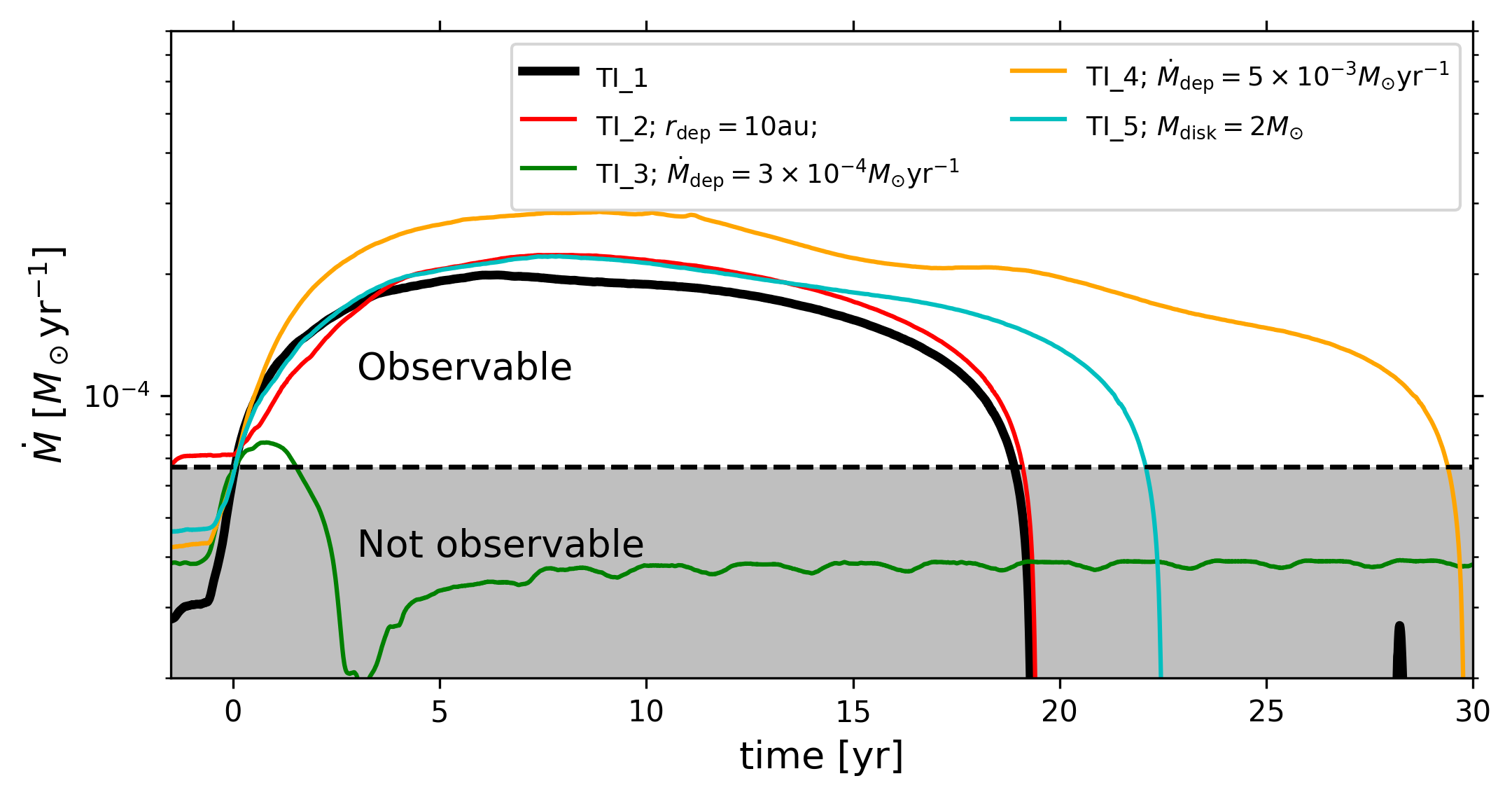}
    \caption{Mass accretion rate during a single TI outburst in various models. The thick black line shows the outburst in our fiducial model. The properties of the models are listed in Table~\ref{tab:1}.}
    \label{fig:arate_multi}
\end{figure}

In this section, we make an attempt to understand how the model parameters are impacting the main properties of the TI outbursts. 

\subsection{\RK{Effect of mass deposition rate and radius}} \label{sec:mdep}
For this, we first vary the values of $r_{\rm dep}$ and $\dot M_{\rm dep}$ (model~\textit{TI\_2}, \textit{TI\_3}, and \textit{TI\_4}). In Figure~\ref{fig:arate_multi} with the thick black line we show the accretion rate history of our fiducial model presented earlier in Section~\ref{sec:TI}. In model \textit{TI\_2} we imply $r_{\rm dep}=10$~au, and it is shown in Figure~\ref{fig:arate_multi} with the red line. Clearly, the value of $r_{\rm dep}$ has almost no influence on the main properties of TI outburst. Next, we vary the value of $\dot M_{\rm dep}$ in our models, assuming $\dot M_{\rm dep}=3\times10^{-4}\msunyr$ in model~\textit{TI\_3} (green line) and $\dot M_{\rm dep}=5\times10^{-3}\msunyr$ in model~\textit{TI\_4} (orange line). Due to the lower $\dot M_{\rm dep}$ in model~\textit{TI\_3}, the TI outburst has very low $\dot M_{\rm peak}\approx8\times10^{-5}\msunyr$, which makes the outburst barely \RK{observable}.

On the other hand, the high $\dot M_{\rm dep}$ in model~\textit{TI\_4} leads to a slight increase in both $\dot M_{\rm peak}$ and the burst duration. However, these values still fall within the range of TI outbursts observed in the fiducial model (see bottom panel in Figure~\ref{fig:mdot_all}). These bursts are comparable in both peak accretion rate and duration because both factors are largely governed by the viscous timescale at the outer edge of the thermally unstable disk region, which participates in the outburst. The latter can be estimated analytically for the steady-state disk as \citep{1994BellLin, 2024Nayakshin}

\begin{equation}
    r_{\mathrm{TI}} = 20\rsun\left(\frac{\dot{M}_{\mathrm{dep}}}{3\times10^{-6}\msunyr}\right)^{1/3} \left( \frac{M_*}{\msun} \right)^{1/3}  \left( \frac{T_{\mathrm{eff}}}{2000} \right)^{-4/3}
\end{equation}

In our models, the region of the disk undergoing TI varies slightly, but since most of the disk mass is located in the outer regions, the duration and peak accretion rates of the bursts remain fairly consistent across models.

In model~\textit{TI\_5} (cyan line), we increase the initial disk mass by a factor of 20, bringing it to 2$\msun$, and include self-gravity in the model. To extend the disk’s lifetime, we also set $\dot M_{\rm dep}=3\times10^{-4}\msunyr$. Despite these changes, the peak accretion rate and the duration of the TI burst increase only slightly compared to the fiducial model. This is again because only the inner $\sim$1~au region of the disk is involved in the TI burst, while the majority of the disk mass remains at larger radii, resulting in minimal impact on the outburst properties.

\subsection{Effect of inner disk radius} \label{sec:rin}
\begin{figure}
    \centering
    \includegraphics[width=1\columnwidth]{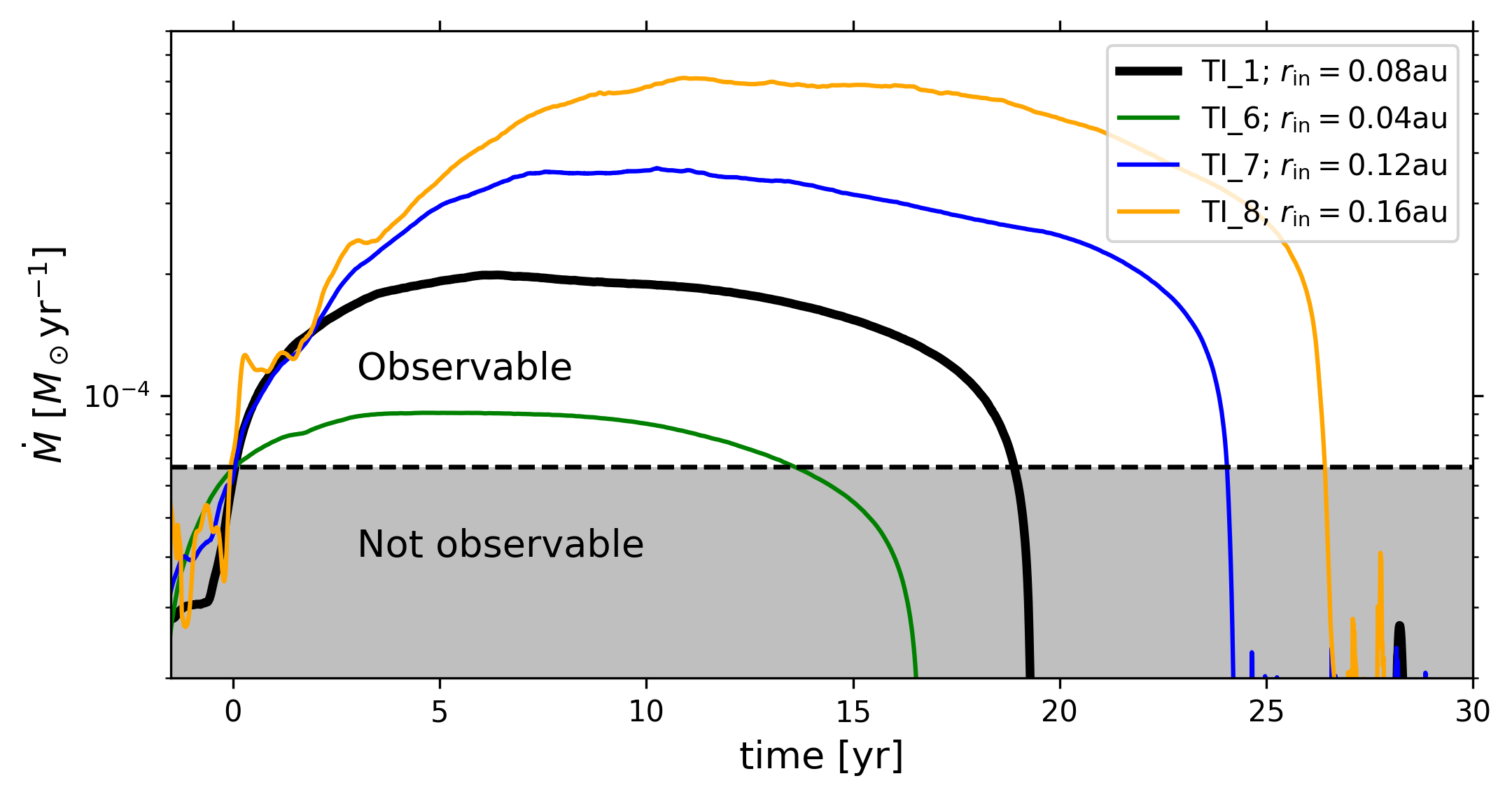}
    \caption{Mass accretion rate during a single TI outburst in models with various $r_{\rm in}$. The thick black line shows the outburst in our fiducial model. The properties of the models are listed in Table~\ref{tab:1}.}
    \label{fig:arate_multi_rin}
\end{figure}

\RK{Contrary}, the peak accretion rate is significantly impacted by changes in the inner disk radius, $r_{\rm in}$. In Figure~\ref{fig:arate_multi_rin}, we compare model \textit{TI\_1}, where $r_{\rm in}=0.08$~au (thick black line), with three other models featuring different $r_{\rm in}$ values: model~\textit{TI\_6} with $r_{\rm in}=0.04$~au (green line), model \textit{TI\_7} with $r_{\rm in}=0.12$~au (blue line), and model \textit{TI\_8} with $r_{\rm in}=0.16$~au (orange line). It is evident that both the duration and peak accretion rate of the TI bursts increase as $r_{\rm in}$ grows larger.

By cutting off a larger portion of the inner disk, we force the TI burst to be triggered farther out. The critical effective temperature for TI bursts depends only weakly on radial distance, but the critical surface density—and therefore the peak mass accretion rate—varies significantly with radius \citep{1994BellLin, 2024Nayakshin}. As a result, as the triggering location moves outward, the S-curve shifts to higher surface densities \citep[cf. fig.~3 in][]{1994BellLin} or \citep[cf. fig.~1 in][]{2024Nayakshin}, leading to a brighter and longer outburst.

For example, in model~\textit{TI\_8}, the TI burst lasts about 26.5 years with a peak accretion rate of $\dot{M}_{\rm peak}\approx6\times10^{-4}\msunyr$, while in model~\textit{TI\_6}, with an inner disk radius four times smaller, the burst lasts about 13.5 years and reaches a peak accretion rate of $\dot{M}_{\rm peak}\approx9\times10^{-5}\msunyr$. 

From these simulations, we find that the peak accretion rate scales approximately as $r_{\rm in}^{1.37}$, whereas the burst duration scales as $r_{\rm in}^{0.5}$. \RK{We treat $r_{\rm in}$ as a free parameter, but in actual high-mass systems the disk inner edge may be set by processes like magnetospheric truncation or magnetic braking \citep[e.g.,][]{2023OlivaKuiper}.} Physically, the stronger dependence of $\dot{M}_{\rm peak}$ on $r_{\rm in}$ likely reflects the increasing mass reservoir and higher disk temperatures in regions closer to the star, while the weaker dependence of the burst duration suggests that local viscous ($\tau_{\rm visc}=r_{\rm TI}^2/\nu$) or thermal timescales grow more slowly with radius. For larger values of $r_{\rm in}$, the TI burst is absent because the disk no longer satisfies the conditions to trigger the thermal instability at these radial distances. Therefore, the choice of $r_{\rm in}$ has a significant impact on the characteristics of TI bursts.

Our simulations yield a scaling of the peak accretion rate with the inner disk radius as $r_{\rm in}^{1.37}$, which is notably shallower than the power laws reported in several 1D steady-state studies — 2.425 \citep{2024Nayakshin}, 2.67 \citep{1998Hameury}, and 2.6 \citep{2004LodatoClarke}. This difference likely stems from the inherent limitations of 1D models, which treat the disk as a vertically averaged, steady-state system and thus concentrate energy release and mass transport in a single layer, resulting in a steeper dependence on $r_{\rm in}$. In contrast, our 2D models resolve the vertical structure, allowing energy to be distributed more gradually throughout the disk. This additional degree of freedom reduces the sensitivity of the peak accretion rate to changes in $r_{\rm in}$, resulting in a lower scaling exponent. Moreover, our approach incorporates non-steady state dynamics and the complex interplay between radial and vertical flows—effects that are absent in 1D models—leading to a more realistic description of thermal instability in accretion disks.

\subsection{Impact of viscosity parameters} \label{sec:alpha}
\begin{figure}
    \centering
    \includegraphics[width=1\columnwidth]{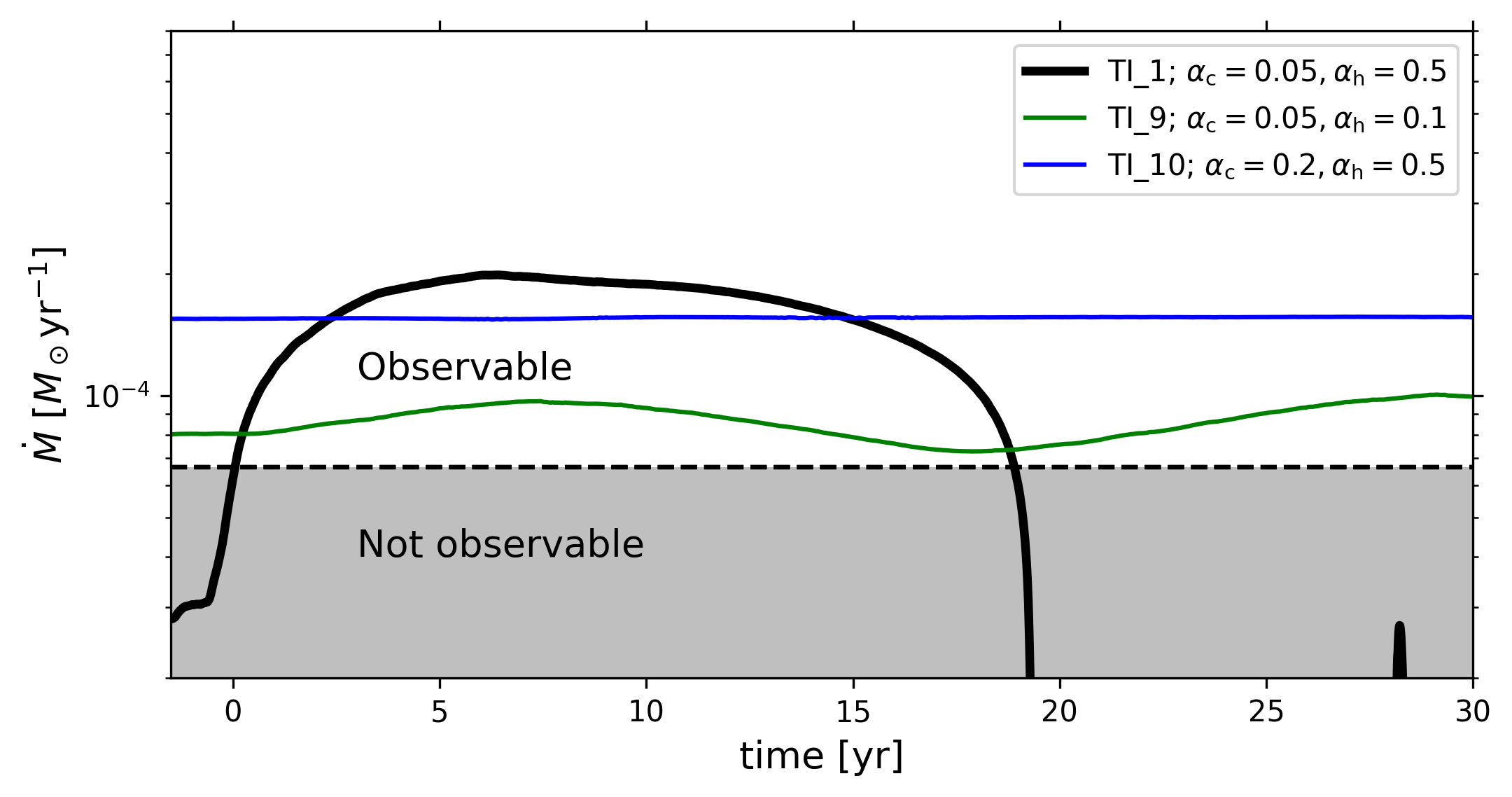}
    \caption{Mass accretion rate during a single TI outburst in models with various $\alpha_{\rm cold}$ and $\alpha_{\rm hot}$ parameters. The thick black line shows the outburst in our fiducial model. The properties of the models are listed in Table~\ref{tab:1}.}
    \label{fig:arate_multi_alpha}
\end{figure}

Continuing from our previous discussion on how the inner disk radius affects TI burst properties, we now turn to another crucial parameter: the viscous $\alpha$ values. The values of $\alpha_{\rm cold}$ and $\alpha_{\rm hot}$ play a pivotal role in shaping the S-curve, which governs the onset and characteristics of thermal instability outbursts. Their relative values determine the stability of the disk and directly influence the amplitude, duration, and even the occurrence of TI bursts.

To investigate the impact of these parameters, we modified model~\textit{TI\_1} (with $\alpha_{\rm cold}=0.05$ and $\alpha_{\rm hot}=0.5$) \RK{by varying the individual $\alpha$ values}. In Figure~\ref{fig:arate_multi_alpha}, we compare the outburst in model~\textit{TI\_1} with model~\textit{TI\_9} \RK{(green line; $\alpha_{\rm cold}=0.05$, $\alpha_{\rm hot}=0.1$)} and model~\textit{TI\_10} \RK{(blue line; $\alpha_{\rm cold}=0.2$, $\alpha_{\rm hot}=0.5$)}. In model~\textit{TI\_9}, the reduced hot-state viscosity ($\alpha_{\rm hot}=0.1$) leads only to low-amplitude variability (less than a factor of 1.5) rather than full TI outbursts. Conversely, in model~\textit{TI\_10}, raising the cold-state viscosity to $\alpha_{\rm cold}=0.2$ keeps the \REF{inner, thermally unstable region of the} disk permanently on the hot branch, suppressing TI bursts entirely. We also tested $\alpha_{\rm cold}=0.01$ and $\alpha_{\rm hot}=0.5$; here the very low cold-state viscosity locks the disk on the cold branch, again preventing any TI activity.

These results underscore the sensitivity of TI burst behavior to the chosen viscous parameters. Future work could further explore the parameter space and investigate how these variations influence observable signatures, potentially providing a diagnostic tool for understanding disk viscosity in high-mass star formation.

\subsection{Role of spatial resolution} \label{sec:resolution}
\begin{figure}
    \centering
    \includegraphics[width=1\columnwidth]{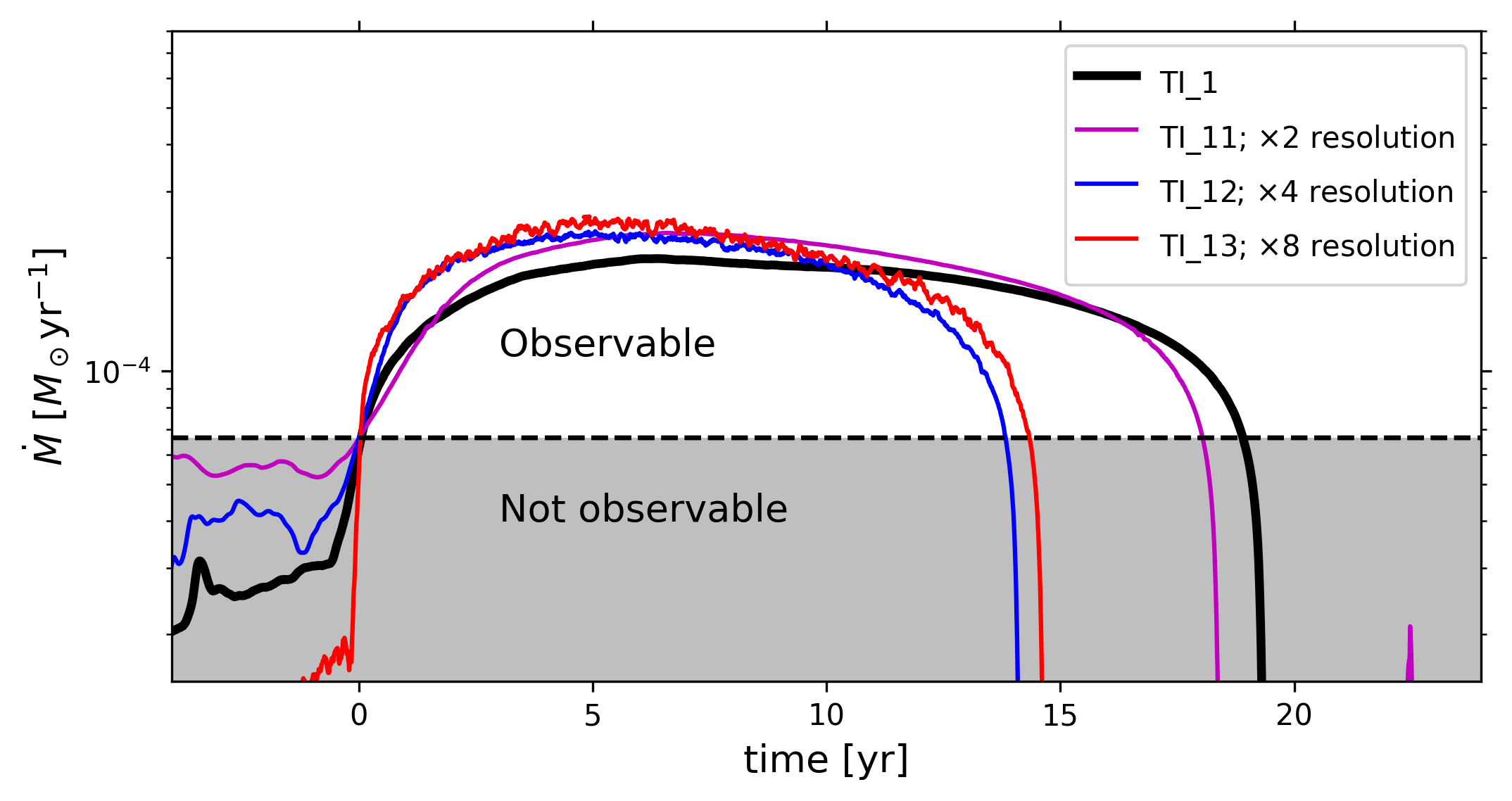}
    \caption{Mass accretion rate during a single TI outburst in models with various spatial grid resolutions. The thick black line shows the outburst in our fiducial model. The properties of the models are listed in Table~\ref{tab:1}.}
    \label{fig:arate_multi_res}
\end{figure}

We also investigated the influence of spatial resolution on the properties of TI bursts in our models by running our fiducial model~\textit{TI\_1} at 2$\times$, 4$\times$, and 8$\times$ higher resolution. In Figure~\ref{fig:arate_multi_res}, we compare a single TI outburst from the higher resolution models with that from our fiducial model. Doubling the resolution (model~\textit{TI\_11}) has almost no impact on the burst duration, with only minor adjustments observed. However, in model~\textit{TI\_12}, where the resolution is increased by a factor of four, the burst duration is approximately 25\% shorter — about 14 years — and the peak accretion rate increases by roughly 15\%, reaching $2.35\times10^{-4}\msunyr$. \ACG{This behavior likely arises because a finer grid better resolves temperature and density gradients \RK{of the vertical stratification}, especially in the inner disk. As a result, the outburst can initiate and develop slightly differently, leading to a shorter duration and a modestly higher accretion peak.} Although the duration in model~\textit{TI\_12} is slightly lower than the range of 15--30 years seen in our fiducial \textit{TI\_1} model (see the bottom panel of Figure~\ref{fig:mdot_all}), the peak accretion rate remains well within the range established by the fiducial model.

\ACG{An additional aspect is the variation in the pre-burst accretion rate among different resolutions. 
These differences reflect the non-steady, evolving nature of the disk rather than a simple resolution artifact. Because the system is not in a perfect steady state, small changes in resolution can alter how mass flows and how instabilities develop before the burst. Moreover, the pre-burst accretion rates — on the order of a few $\times 10^{-5}\,\msunyr$ — are typically overshadowed by the star’s luminosity, making them unlikely to be observed directly.}

\REF{To check that our bursts are well resolved vertically, we measured the disk aspect ratio $h=H/r$ and the number of polar cells per scale height both before and during the burst. In the fiducial model~\textit{TI\_1}, during quiescence $h\approx0.032$ at 0.1~au (2 cells per $H$) and $0.021$ at 0.3~au (1 cell per $H$), while at burst peak $h$ grows to 0.167 (8 cells per $H$ at 0.1~au) and 0.191 (9 cells per $H$ at 0.3~au). In the highest‑resolution model~\textit{TI\_13}, even the quiescent disk is well resolved (6–8 cells per $H$), and the puffed‑up disk at burst peak spans 32 cells per $H$ at both 0.1~au and 0.3~au. The close agreement in burst duration and peak accretion rate between these two extremes confirms that our results have effectively converged in the vertical direction.}

Despite these variations, our highest-resolution model (\textit{TI\_13}) shows similar burst properties to \textit{TI\_12}, suggesting that the outburst duration \RK{might be} converged. Overall, while increasing grid resolution can introduce modest shifts in both pre-burst and burst-phase accretion behavior, the main conclusions remain robust: peak accretion rates consistently fall in the range of $2-3\times10^{-4}\,\msunyr$, and burst durations exceed 10 years. This indicates that our models already capture the essential physics of TI bursts and that further increases in resolution are unlikely to significantly alter these key results.

\subsection{Dependence on stellar mass} \label{sec:st_mass}
\begin{figure}
    \centering
    \includegraphics[width=1\columnwidth]{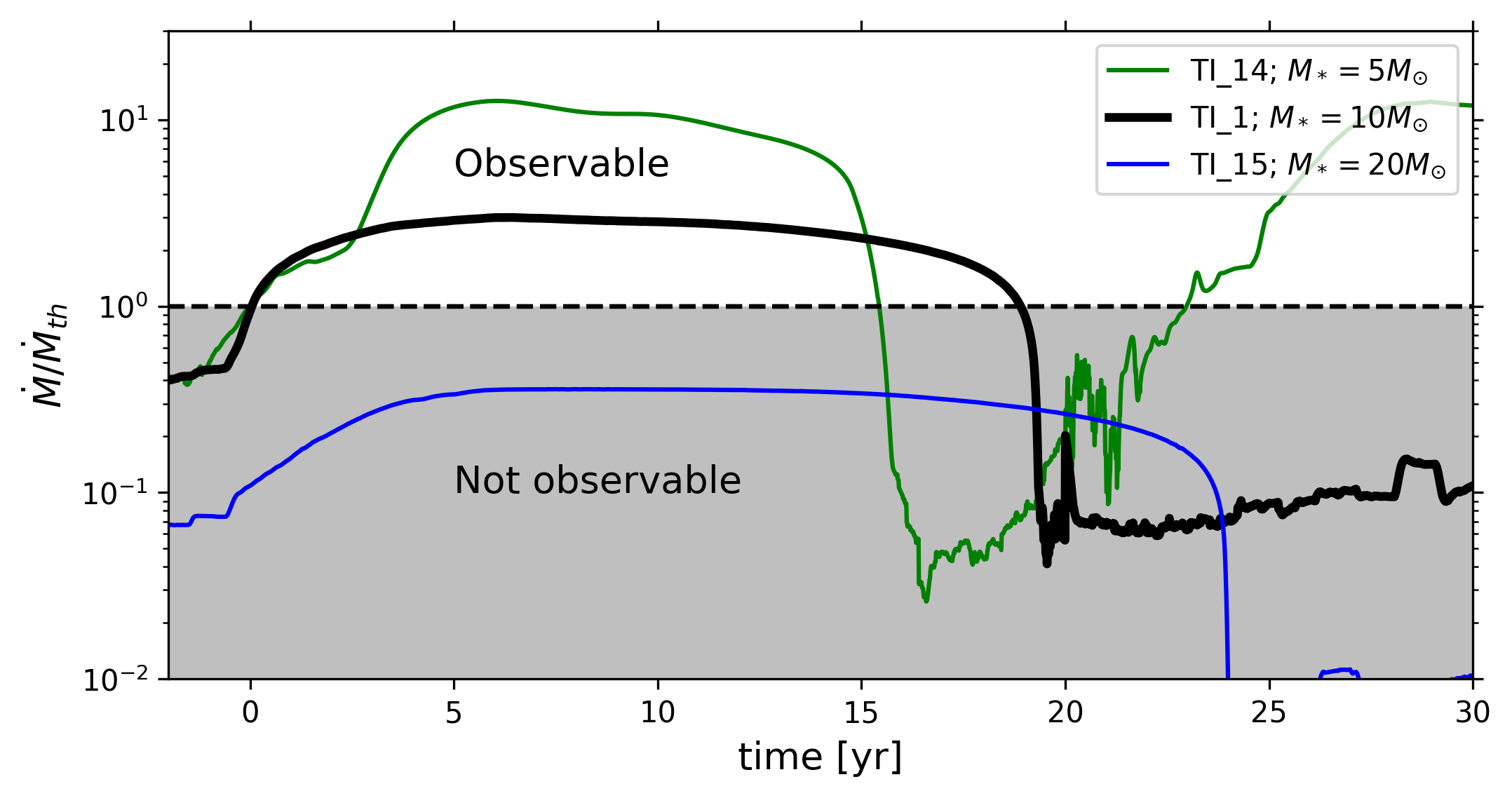}
    \caption{Mass accretion rate, \RK{normalized to $\dot{M}_{\rm th}$ for each stellar mass,} during a single TI outburst in models with various stellar masses. The thick black line shows the outburst in our fiducial model. The properties of the models are listed in Table~\ref{tab:1}.}
    \label{fig:arate_multi_mass}
\end{figure}

So far, we have discussed TI outbursts in HMYSOs assuming a stellar mass of $M_*$=$10\msun$. However, observed outbursts occur in HMYSOs with stellar masses ranging from approximately $5\msun$ to a few tens of $\msun$ (\ACG{e.g.} see Table~3 in \citet{2024Elbakyan} \ACG{and Table 5 in \citet{2024WolfStecklum}}). To explore how stellar mass affects TI bursts, in Figure~\ref{fig:arate_multi_mass} we compare the TI outburst in model~\textit{TI\_1} with those in model~\textit{TI\_14} (solid green line, $M_*$=$5\msun$) and model~\textit{TI\_15} (solid blue line, $M_*$=$20\msun$). \RK{We normalize the mass accretion rates by the corresponding threshold accretion rates, $\dot{M}_{\rm th}$, for each stellar mass.} 

To ensure a physically consistent comparison, we adjusted the inner disk radius based on stellar mass: in model~\textit{TI\_14}, $r_{\rm in}$=$0.04$ au for the $ 5\msun$ star, while in model~\textit{TI\_15}, $ r_{\rm in}$=$0.12 $ au for the $ 20\msun $ star. 
\RK{This choice is reasonable under the simplifying assumption that the disk’s truncation radius scales roughly with the stellar radius — expected to increase for higher-mass stars — although real inner disk physics is more complex, involving processes such as magnetic braking, radiation pressure, and boundary‐layer effects.}

The TI outburst in model~\textit{TI\_14} lasts approximately 15 years, with a peak accretion rate of $\dot{M}_{\rm peak} \approx 9.2 \times 10^{-5} \msunyr$. This is more than an order of magnitude lower than the accretion rates observed in YSOs of similar mass \citep{2021Chen, 2021Hunter}. Additionally, we note that the outburst frequency in model~\textit{TI\_14} increases, with outbursts occurring periodically on timescales of less than 10 years. In model~\textit{TI\_15}, the peak accretion rate is $\dot{M}_{\rm peak} \approx 3.8 \times 10^{-5} \msunyr$, which is below the threshold value of $\dot{M}_{\rm th} = 6.1 \times 10^{-5} \msunyr$ for that stellar mass. This means that such an outburst would not be observable.  

\ACG{It is important to note that in these comparisons we have varied only the stellar mass and the inner disk radius while keeping the disk mass, mass deposition rate ($\dot{M}_{\rm dep}$), and deposition radius ($r_{\rm dep}$) fixed. In a fully self-consistent model, one would expect these parameters to scale with the stellar mass as well—for instance, more massive systems would naturally have higher disk masses and accretion rates. Our approach isolates the impact of varying $M_*$ and $r_{\rm in}$ to explore their individual effects on TI bursts, but we acknowledge that this is not entirely physical. Preliminary tests where disk mass and $\dot{M}_{\rm dep}$ were varied indicate that while such changes can alter the burst amplitude and duration, the overall trends remain similar. A more complete investigation of parameter rescaling in massive systems will be addressed in future work.}

\section{\REF{Disk mass transport and accumulation}} \label{sec:app_disk_mass}

\begin{figure}
    \centering
    \includegraphics[width=1\columnwidth]{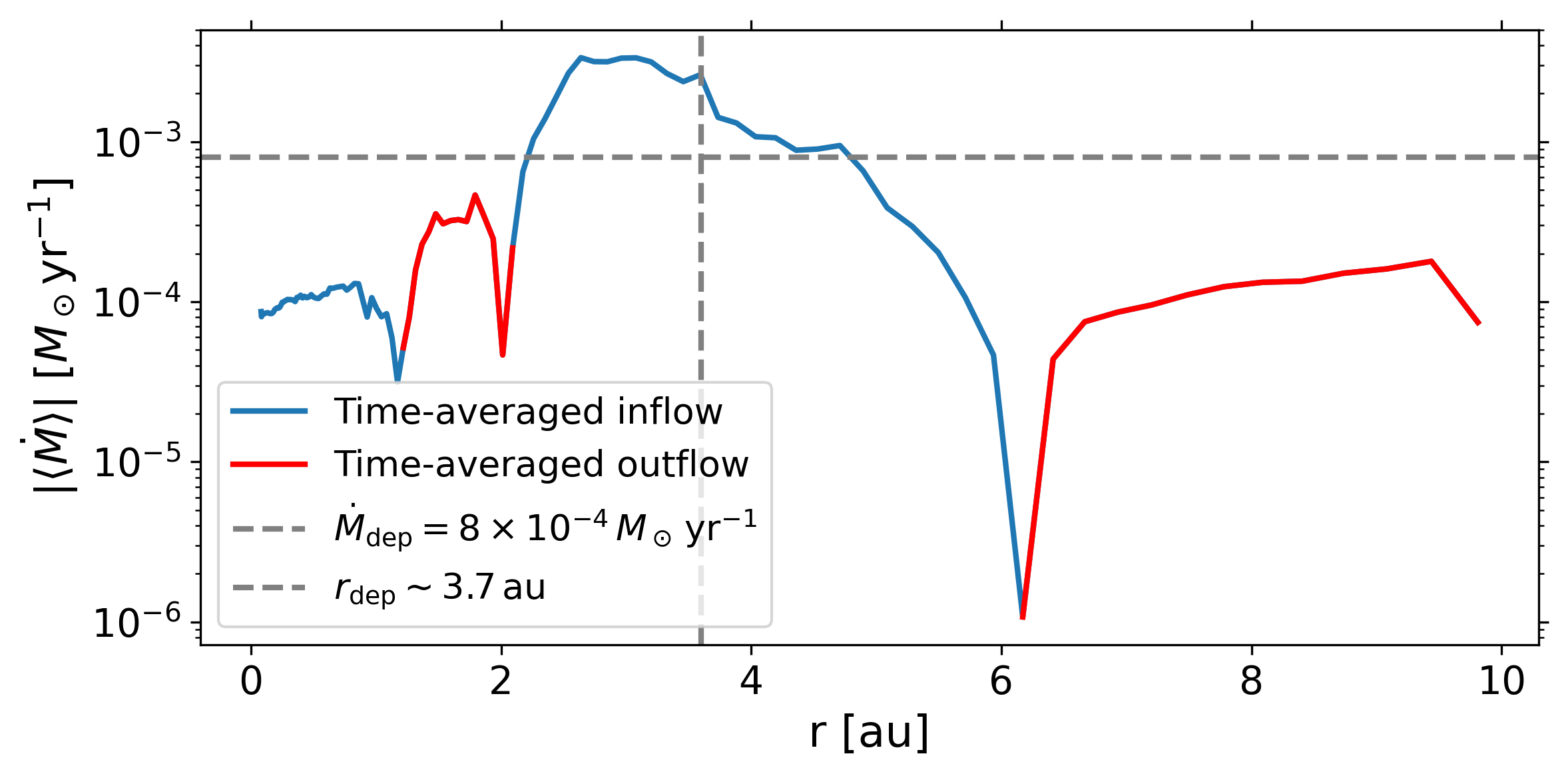}
    \caption{\REF{Time-averaged radial mass transport rate for the fiducial model~\textit{TI\_1}. The blue curve shows the absolute value of the time-averaged mass transport rate inwards, while the red curve for outwards. The horizontal dashed line marks the constant deposition rate, $\dot M_{\rm dep}=8\times10^{-4}\msunyr$ and the vertical dashed line indicates the deposition radius $r_{\rm dep} \sim 3.7$~au. Mass transport rates are averaged over the simulation time interval of 1000 years shown in Fig.~\ref{fig:mdot_all}.}}
    \label{fig:time_aver_mdot}
\end{figure}

\REF{
We measured the net mass flux through both the inner and outer radial boundaries and found that, after an initial adjustment of roughly 2000~yr, both settle to nearly constant values of order $\sim10^{-4}\msunyr$.  These are well below our imposed deposition rate of $\dot M_{\rm dep}=8\times10^{-4}\msunyr$, so the disk as a whole continues to gain mass, especially near the deposition radius $r_{\rm dep}$.
Fig.~\ref{fig:time_aver_mdot} shows the time‑averaged radial mass transport for our fiducial model~\textit{TI\_1}.  Inside $r_{\rm dep}$, the region from about 1~au to 2~au actually exhibits a net outward flow, which results from the strong TI bursts in the inner disk (see Sect.~\ref{sec:TI} and Fig.~\ref{fig:mdot_rad}), whose explosive redistribution pushes material outward and temporarily impedes the inward drift.  Closer in, at $r\lesssim1$~au, the mean transport remains directed inward but at a rate much lower than $\dot M_{\rm dep}$.
Outside $r_{\rm dep}$, material again flows inward between $\sim r_{\rm dep}$ and $\sim6$~au, while at larger radii ($r\gtrsim6$~au) viscous spreading drives a weak outward flux.  Since both the inner inflow and outer spreading rates are far too small to balance the deposition, mass gradually accumulates across the disk.  Although the boundary fluxes themselves reach a quasi‑steady value, true steady state, where the disk mass no longer changes, would require those fluxes to match $\dot M_{\rm dep}$, which they do not.
}

\begin{figure}
    \centering
    \includegraphics[width=1\columnwidth]{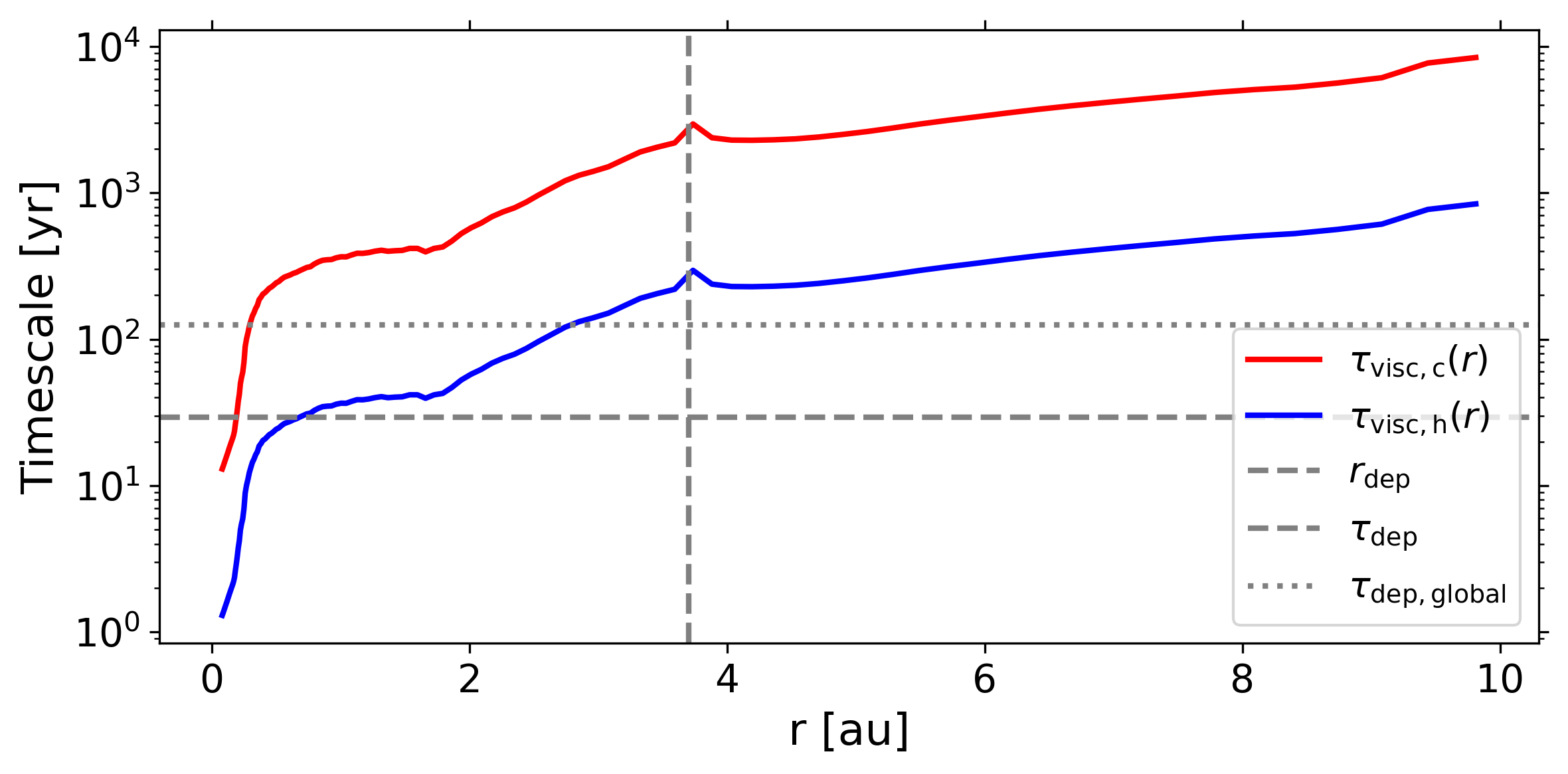}
    \caption{\REF{Radial profiles of characteristic timescales in the disk. The red and blue curves show the time-averaged viscous timescales for $\alpha=\alpha_{\rm c}=0.05$ and $\alpha=\alpha_{\rm h}=0.5$ disk regions, respectively. 
    The horizontal dashed line marks the mean mass-deposition timescale $\tau_{\rm dep}$ at the characteristic deposition radius $r_{\rm dep}$ (vertical dashed line). The horizontal dotted line marks the global mass‑deposition timescale, $\tau_{\rm dep,global}$=$M_{\rm disk}(t=0)/\dot M_{\rm dep}\approx125$~yr. }}
    \label{fig:timescales}
\end{figure}

\REF{
We have also evaluated three key timescales (the mass deposition timescale, and the viscous timescales in both the cold and hot regimes) to understand why mass accumulates rather than draining through the disk. In Fig.~\ref{fig:timescales} the viscous timescale in quiescence, $\tau_{\rm visc,c}(r) = r^2/\nu_{\rm c}\quad(\alpha_{\rm c}=0.05)$, is shown by the red curve, and the viscous timescale during an outburst, $\tau_{\rm visc,h}(r) = r^2/\nu_{\rm h}\quad(\alpha_{\rm h}=0.5)$, by the blue curve.  These reflect how quickly viscous transport can redistribute mass under “cold” and “hot” effective viscosities, respectively.}
\REF{
To compare with our imposed additional mass infall, we define a mass‑deposition timescale at $r_{\rm dep}$,
\begin{equation} 
    \tau_{\rm dep}\;=\;\frac{\Sigma(r_{\rm dep})\,2\pi r_{\rm dep}\,\Delta r}{\dot M_{\rm dep}},
\end{equation}
plotted as the horizontal dashed line. Physically, $\tau_{\rm dep}$ is the time needed to fill the annulus at $r_{\rm dep}$ by our constant deposition rate. Here $\Delta r\approx0.15$~au is the radial width of the grid cell at $r_{\rm dep}$ where mass is injected. Although this choice is somewhat arbitrary (cell sizes in our logarithmic mesh range from $\sim$0.003~au to 0.38~au) taking a smaller $\Delta r$ would only shorten $\tau_{\rm dep}$ further. Even if we instead used the entire initial disk mass, $M_{\rm disk}(t=0)=0.1\msun$, the global deposition timescale $M_{\rm disk}/\dot M_{\rm dep}\approx125$~yr remains well below the viscous timescales. Because $\tau_{\rm dep}$ is about 4 times shorter than $\tau_{\rm visc,h}$ and about 40 times shorter than $\tau_{\rm visc,c}$ at that location, mass arrives much faster than the disk can redistribute it, even in the hot state.}
\REF{
This imbalance explains the steady buildup of disk mass: the inflow outpaces viscous transport for most of the disk, so the disk continually grows until a burst occurs.}
Over longer times, the outer mass flux may become comparable to the deposition rate and establish a time averaged steady state, but our simulations have not yet reached this regime. If the disk grows sufficiently massive, gravitational instability could also drive episodic mass loss until balance is restored. We will explore the longer term evolution and self gravity effects in future studies.

\begin{figure}
    \centering
    \includegraphics[width=1\columnwidth]{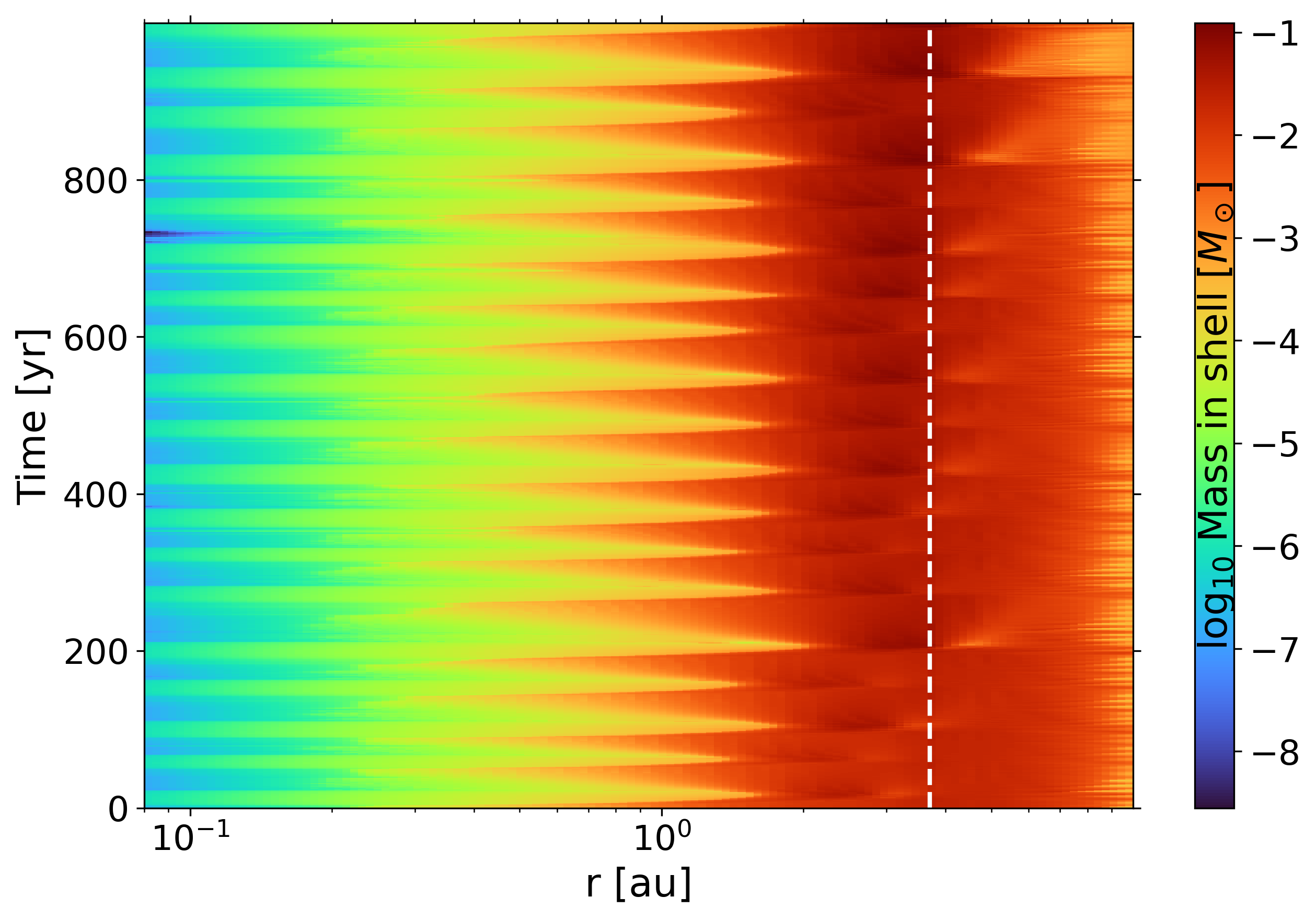}
    \caption{\REF{Time evolution of the radial mass distribution in the computational domain. The color scale shows the logarithm of the mass contained in each radial shell. The vertical dashed line marks $r = 3.7$~au.}}
    \label{fig:disk_mass_time}
\end{figure}

\REF{We note that the mass is deposited uniformly into the midplane at $r_{\rm dep}\approx3.7$~au. In a logarithmic radial grid, the physical width of each cell increases with radius. Thus, a modest increase in the cell density at the deposition radius (e.g., left panels in Fig.~\ref{fig:dens_temp_2d}) actually corresponds to a large local mass injection when integrated over the cell’s full volume. Even if the color map shows only a slightly brighter band at $r_{\rm dep}$, that cell contains much more mass than an identically bright cell closer in, simply because of its larger volume. In other words, the eye can underestimate how much material piles up at the injection radius.}

\REF{To make this accumulation clear, Figure~\ref{fig:disk_mass_time} plots the time evolution of the radial mass profile of the disk.  The vertical dashed line marks $r_{\rm dep}$, and one sees that the added material indeed piles up close to $r_{\rm dep}$.  Between bursts, the inner disk gradually refills, then is rapidly drained during each TI outburst, producing the characteristic sawtooth pattern in mass versus time.}

\REF{Our simulations reveal that during TI bursts the inner disk becomes significantly “puffed up”.  In such a thick disk, one might expect deviations from purely Keplerian rotation to suppress the usual $\alpha P$ viscous torque and thereby reduce mass transport onto the star.  To quantify this effect, we compared two measures of the viscous torque $G(r)$ in our fiducial model~\textit{TI\_1}: the standard “Keplerian” torque,
\begin{equation}
    G_{\rm K}(r) \;=\; 2\pi\,r^2\,\alpha\,P(r)\,\bigl|\partial_r\Omega_{\rm K}\bigr|,
\end{equation}
and the “actual” torque computed with the local shear,
\begin{equation}
    G_{\rm act}(r) \;=\; 2\pi\,r^2\,\alpha\,P(r)\,\bigl|\partial_r(v_\phi/r)\bigr|,
\end{equation}
where $v_\phi$ is the midplane azimuthal velocity.}

\REF{We find that although the disk remains nearly Keplerian, with the midplane azimuthal velocity deviating by less than 1\% in quiescence and up to 7–10\% at burst peak, the viscous torque, which scales with the radial gradient of $v_\phi/r$, is suppressed by roughly 30\% during quiescence and 40–45\% during the outburst. In our fiducial model, the peak accretion rate onto the star $\dot{M}_{\rm peak}$ during a burst is about a factor of four lower than the mass deposition rate $\dot{M}_{\rm dep}$. While the non‐Keplerian reduction in viscous torque accounts for a substantial fraction of this deficit, it is not sufficient on its own to explain the entire gap between infall and accretion. Consequently, additional effects, such as the burst‐driven convective and outflowing motions, must also play key roles in regulating the net mass transport.}

\section{\REF{S-curves in 1D and 2D models}} \label{sec:s_curves}

\begin{figure}
    \centering
    \includegraphics[width=1\columnwidth]{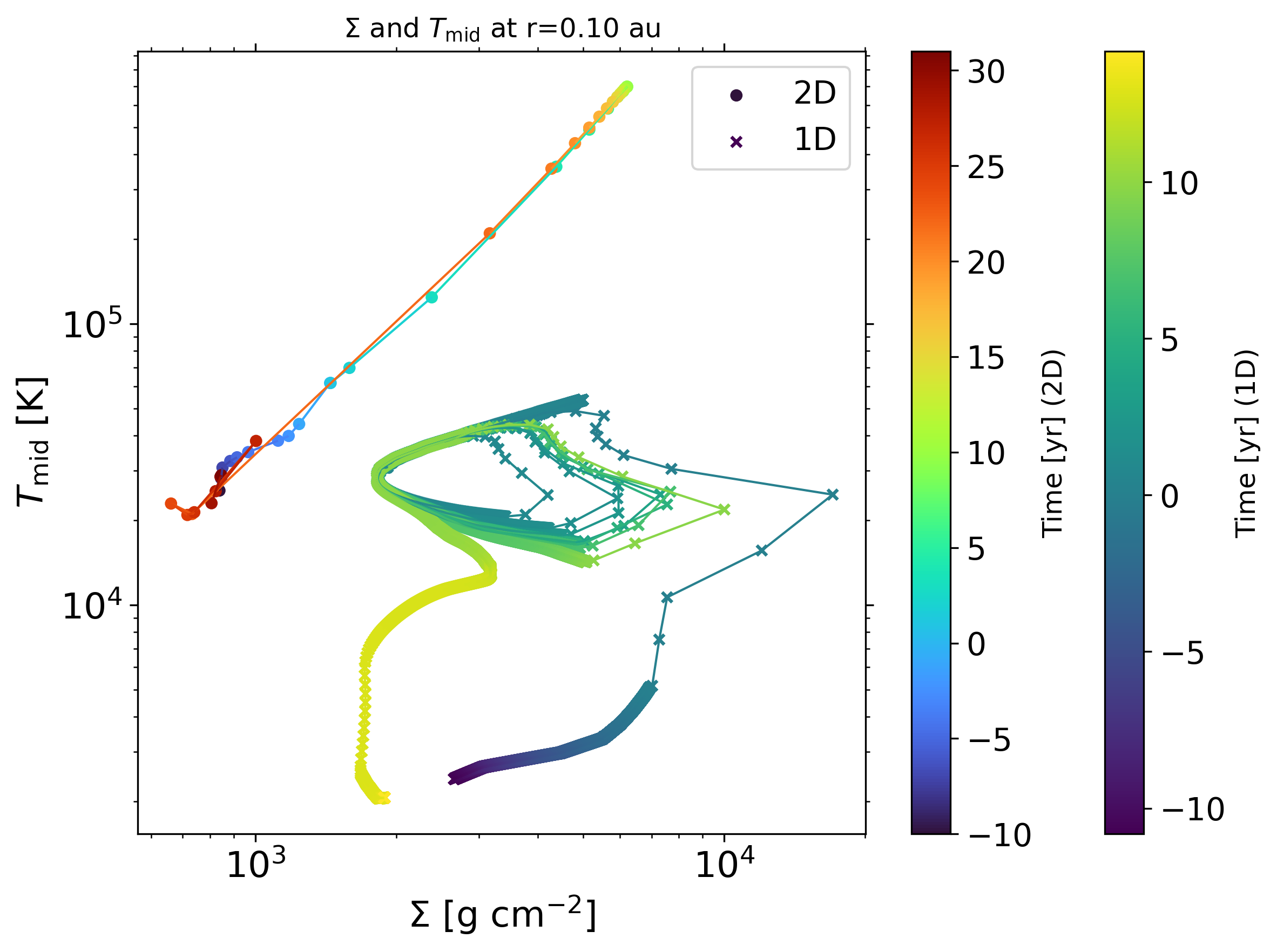}
    \includegraphics[width=1\columnwidth]{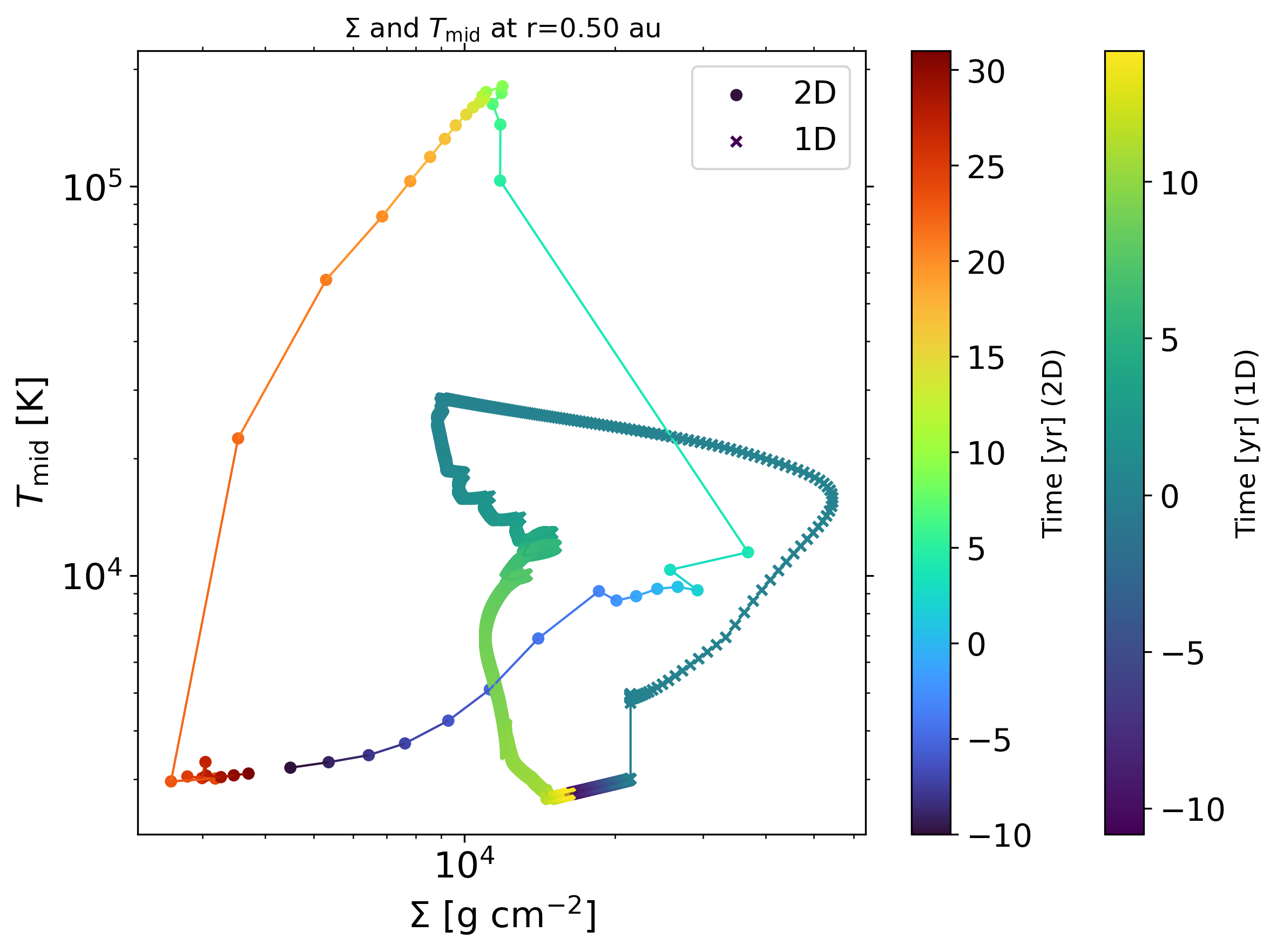}
    \includegraphics[width=1\columnwidth]{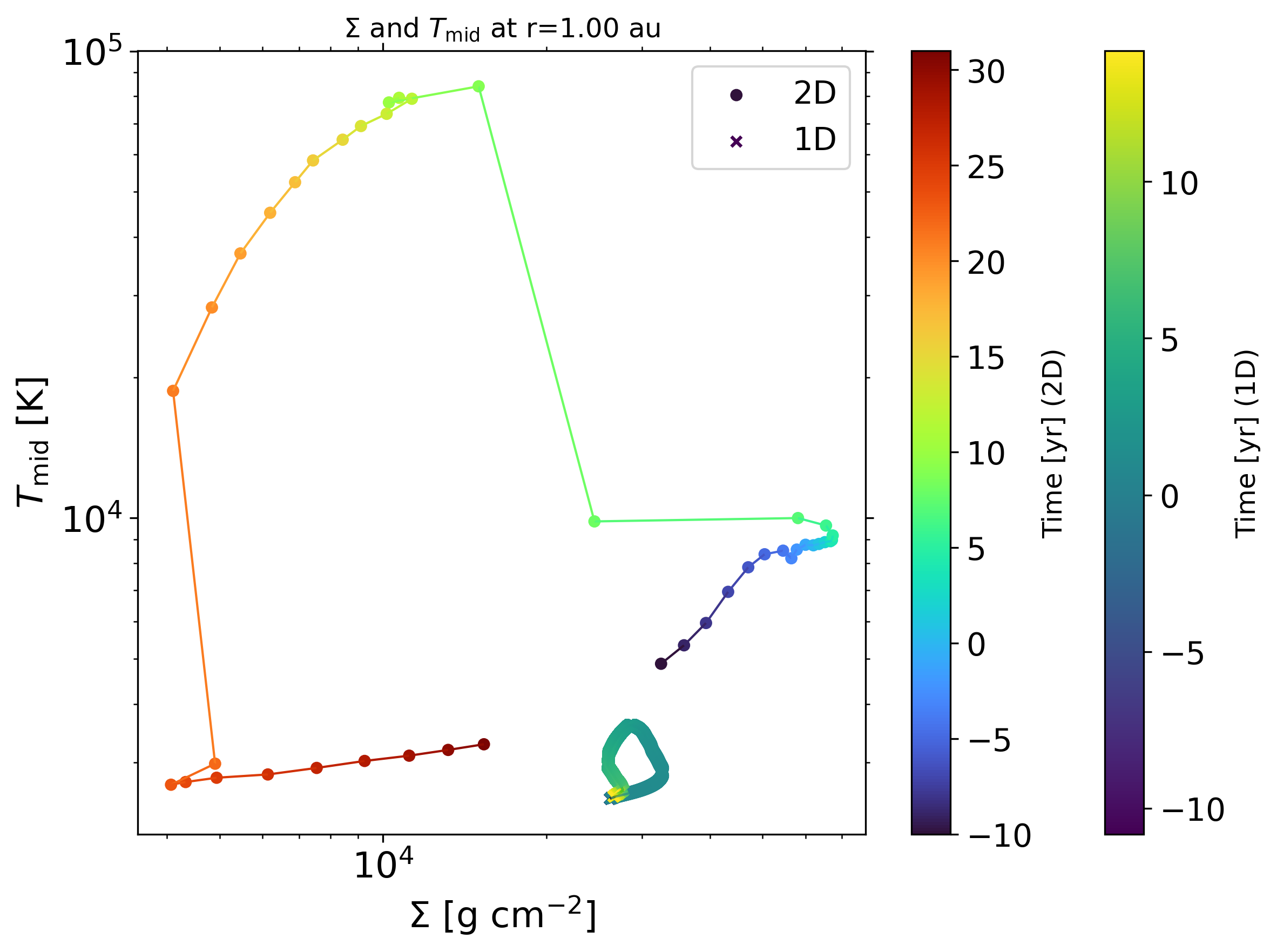}
    \caption{\REF{Time evolution of the midplane temperature $T_{\rm mid}$ versus surface density $\Sigma$ at $r=0.1$~au (top panel), $r=0.5$~au (middle panel), and $r=1.0$~au (bottom panel) for our 1D \citep{2024Elbakyan} and 2D (this work, model~\textit{TI\_1}) TI models. Colored points and lines show the trajectory from the 2D model. Crosses and lines show the corresponding evolution from the 1D model. Time $t=0$ corresponds to the beginning of the TI burst.
    }}
    \label{fig:s_curve}
\end{figure}

\REF{To illustrate how the inclusion of vertical structure alters the TI cycle, we compare the evolution of the inner disk in our 1D \citep{2024Elbakyan} and 2D (this work, model~\textit{TI\_1}) simulations in the plane of surface density ($\Sigma$) versus midplane temperature ($T_{\rm mid}$). 
We note that although the key parameters (stellar mass, disk mass, viscosity coefficients, critical temperature) are comparable between the models, the 1D run adopts a lower mass deposition rate of $10^{-4}\msunyr$ (following \citet{2024Elbakyan}), while our 2D fiducial model uses $8\times10^{-4}\msunyr$. This choice in the 1D case accentuates certain S‑curve features that tend to be smoothed out at higher infall rates. Furthermore, the two setups differ not only in dimensionality but also in their thermodynamic treatments: our 2D model uses a simplified equation of state with a fixed mean molecular weight, whereas the 1D calculations employ a more detailed equation of state that accounts for hydrogen dissociation and ionization. Our goal is to demonstrate how vertical stratification and non‐steady behavior reshape the classical thermal‐instability cycle. Figure~\ref{fig:s_curve} shows the track of the annulus at $r=0.1$~au (top panel), $r=0.5$~au (middle panel), and $r=1.0$~au (bottom panel) over one TI cycle in both setups, highlighting how the S‐curve “knee” location and loop shape shift with radius.}

\REF{At 0.1~au, the 2D annulus begins at roughly $(\Sigma, T_{\rm mid})\approx(8\times10^{2}$~g~cm$^{-2}$, $2\times10^{4}$~K). Rather than showing a sudden knee, its density and temperature rise together in a smooth, diagonal path all the way to $(6\times10^{3}$~g~cm$^{-2}$, $7\times10^{5}$~K), and then retrace nearly the same route as they cool. In contrast, the 1D curve sits initially at $(2.6\times10^{3}$~g~cm$^{-2}$, $2.5\times10^{3}$~K), warms smoothly to $(1.8\times10^{4}$~g~cm$^{-2}$, $2.5\times10^{4}$~K), then executes a jump to $(5\times10^{3}$~g~cm$^{-2}$, $5\times10^{4}$~K) before shifting along the hot branch and returning close to its start point. Smaller loops on the return track arise from the reflares \citep{2000Menou, 2001Dubus} discussed in Sect.~6.2 of \citep{2024Elbakyan}. This two‐stage 1D behavior produces a brief, intense spike in accretion, while the 2D path yields a much longer, gentler burst. As we noted earlier, differences in the equation of state used may partly explain the difference outlined above; non‑local vertical motions and energy losses almost certainly play a major role as well, and we plan to explore these effects in future work.}

\REF{Moving out to 0.5~au, the 2D loop again unfolds over a wide range of both $\Sigma$ and $T_{\rm mid}$. The annulus departs the cold branch at about $(4.5\times10^{3}$~g~cm$^{-2}$, $3\times10^{3}$~K), climbs steadily to $(3.8\times10^{4}$~g~cm$^{-2}$, $10^{4}$~K), then jumps to $(1.2\times10^{4}$~g~cm$^{-2}$, $1.5\times10^{5}$~K), and finally cools and contracts back toward its origin. In 1D at this radius, however, the loop is much tighter: a small vertical heating jump at $\Sigma\approx2.1\times10^{4}$~g~cm$^{-2}$ is followed by a drift to $(5.5\times10^{4}$~g~cm$^{-2}$, $1.8\times10^{4}$~K) before returning. The dominance of vertical energy losses in 2D means that even at intermediate radii the burst involves significant mass reshuffling during heating, stretching the loop and lengthening the event relative to 1D.}

\REF{At 1.0~au, the 2D trajectory still completes a pronounced loop — heating from $(3.1\times10^{4}$~g~cm$^{-2}$, $5\times10^{3}$~K) up to $(1.5\times10^{4}$~g~cm$^{-2}$, $8\times10^{4}$~K) and then cooling and rebounding. In contrast, the 1D annulus barely leaves the cold branch, executing only a small, circular path near the low‑temperature knee of the S‑curve, as described in \citep[][ Sect.~6.1.3]{1994BellLin}. There, cooler disk regions cannot heat sufficiently to jump to the upper unstable branch and instead sample a localized loop around the cold knee and the lower unstable branch of the S-curve. As discussed by \citet{1994BellLin}, non-local energy transfer is important for this behavior. In our 2D models, enhanced radial and vertical energy transport enlarges the region that participates in the instability, so that even relatively cool annulus at 1~au undergo substantial TI‐driven excursions.}

\REF{In summary, our 2D simulations reveal two key differences from the classic 1D picture. First, at small radii the disk never fully settles back onto the cold branch during an outburst, instead remaining in a partially heated state. Second, the thermally unstable zone extends much farther into what would be the “cold” disk in 1D models.}

\end{appendix}

\end{document}